\documentclass[sigconf]{acmart}

\AtBeginDocument{%
  \providecommand\BibTeX{{%
    \normalfont B\kern-0.5em{\scshape i\kern-0.25em b}\kern-0.8em\TeX}}}

\setcopyright{rightsretained}

\acmSubmissionID{pn9848}

\begin{document}

\copyrightyear{2021}
\acmYear{2021}
\acmConference[CHI '21]{CHI Conference on Human Factors in Computing Systems}{May 8--13, 2021}{Yokohama, Japan}
\acmBooktitle{CHI Conference on Human Factors in Computing Systems (CHI '21), May 8--13, 2021, Yokohama, Japan}\acmDOI{10.1145/3411764.3445748}
\acmISBN{978-1-4503-8096-6/21/05}

\title[Transparency, Equity, and Community Challenges for Student Assignment Algorithms]{Modeling Assumptions Clash with the Real World: Transparency, Equity, and Community Challenges for Student Assignment Algorithms}

\author{Samantha Robertson}
\affiliation{%
  \institution{University of California, Berkeley}
  \city{Berkeley}
  \state{California}
}
\email{samantha_robertson@berkeley.edu}

\author{Tonya Nguyen}
\affiliation{%
  \institution{University of California, Berkeley}
  \city{Berkeley}
  \state{California}
}
\email{tonyanguyen@berkeley.edu}

\author{Niloufar Salehi}
\affiliation{%
  \institution{University of California, Berkeley}
  \city{Berkeley}
  \state{California}
}
\email{nsalehi@berkeley.edu}

\renewcommand{\shortauthors}{Robertson, Nguyen, and Salehi}

\begin{abstract}
Across the United States, a growing number of school districts are turning to matching algorithms to assign students to public schools. The designers of these algorithms aimed to promote values such as transparency, equity, and community in the process. However, school districts have encountered practical challenges in their deployment. In fact, San Francisco Unified School District voted to stop using and completely redesign their student assignment algorithm because it was frustrating for families and it was not promoting educational equity in practice. We analyze this system using a Value Sensitive Design approach and find that one reason values are not met in practice is that the system relies on modeling assumptions about families’ priorities, constraints, and goals that clash with the real world. These assumptions overlook the complex barriers to ideal participation that many families face, particularly because of socioeconomic inequalities. We argue that direct, ongoing engagement with stakeholders is central to aligning algorithmic values with real world conditions. In doing so we must broaden how we evaluate algorithms while recognizing the limitations of purely algorithmic solutions in addressing complex socio-political problems.
\end{abstract}



\begin{teaserfigure}
  \includegraphics[width=\textwidth]{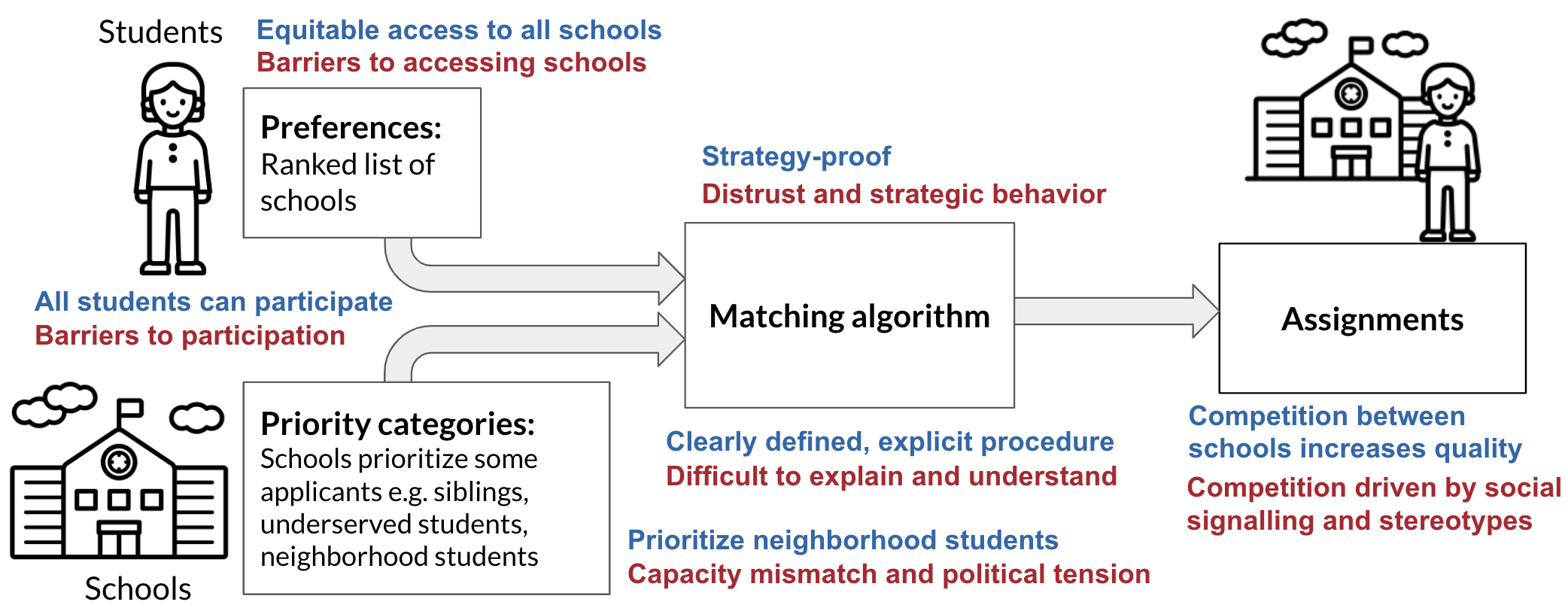}
  \caption{Student assignment algorithms were designed meet school district values based on modeling assumptions (blue/top) that clash with the constraints of the real world (red/bottom). Students are expected to have predefined preferences over all schools, which they report truthfully. The procedure is intended to be easy to explain and optimally satisfies student preferences. In practice however, these assumptions clash with the real world characterized by unequal access to information, resource constraints (e.g. commuting), and distrust.}
   \Description[A flow chart showing the inputs (student preferences and school priorities) to the matching algorithm and its output (student-school assignments), labelled with modeling assumptions and real world challenges at each stage of the process.]{A flow chart showing the inputs and outputs to the matching algorithm. On the left there is a student icon next to a box labelled ``Preferences: Ranked list of schools.'' Below that is a school icon with a box labelled ``Priority categories: Schools prioritize some applicants e.g. siblings, underserved students, neighborhood students.'' These two boxes have arrows leading to a box in the centered labeled ``Matching algorithm.'' This box then leads to a final box on the right labelled ``Assignments'' with an icon showing a student and a school. Surrounding the flow chat are pairs of text showing the modeling assumptions and corresponding real world challenges and constraints. From left to right and top to bottom these read: ``Equitable access to all schools / Barriers to accessing schools;'' ``Strategy-proof / Distrust and strategic behavior;'' ``All students can participate / Barriers to participation;'' ``Clearly defined, explicit procedure / Difficult to explain & understand;'' ``Competition between schools increases quality / Competition driven by social signalling & stereotypes;'' ``Prioritize neighborhood students / Capacity mismatch and political tension.''}
    \label{fig:splash}
\end{teaserfigure}

\maketitle

\section{Introduction}

Algorithmic systems are increasingly involved in high-stakes decision-making such as child welfare \cite{Saxena2020, Brown2019}, credit scoring \cite{Koren2015}, medicine \cite{Ghassemi2019, Obermeyer2019}, and law enforcement \cite{BarryJester2015}. Documented instances of discriminatory algorithmic decision-making \cite{BarryJester2015, Chouldechova2017, Obermeyer2019, Ali2019} and biased system performance \cite{Buolamwini2018, Noble2018, Sweeney2013, Bolukbasi2016} have prompted a growing interest in designing systems that reflect the values and needs of the communities in which they are embedded \cite{Friedman2003, Zhu2018}. However, even when systems are designed to support shared values, they do not always promote those values in practice \cite{Voida2014}. One reason why an algorithmic system may not support values as expected is that \textit{these expectations rely on modeling assumptions about the world that clash with how the world actually works}. In this paper, we examine one such breakdown, the San Francisco Unified School District's student assignment algorithm, to study where and how those clashes occur and to offer paths forward.

San Francisco has a long history of heavily segregated neighborhoods which has resulted in segregated schools when students attend their neighborhood school \cite{sfusd2018}. In 2011, in an effort to promote educational equity and racially integrated classrooms, San Francisco Unified School District joined many cities across the country who were turning to assignment algorithms to determine where students go to school \cite{sfusd2018}. Rather than enrolling in their neighborhood school, students submit their ranked preference list of schools to the district, and the algorithm uses those preferences along with school priorities and capacity constraints to match students to schools. These algorithms have been met with great excitement for their potential to provide more equitable access to public education and give families more flexibility compared to a neighborhood-based assignment system \cite{Kasman2019}. By 2018, however, diversity in schools had instead decreased and parents were frustrated by an opaque and unpredictable process \cite{sfusd2018}. In fact, many schools were now more segregated than the neighborhoods they were in \cite{sfusd2015}. The algorithm had failed to support the values its designers had intended and the San Francisco Board of Education voted for a complete overhaul and redesign of the system \cite{sfusd2018}. 

Following a Value Sensitive Design approach, we ask two central questions: 1) What values were designers and policy-makers hoping this algorithm would support? 2) Why were those values not met in practice? To answer these questions we first analyzed the school district's publicly available policy documents on student assignment and conducted a review of the relevant economics literature where matching algorithms for student assignment have been developed. To answer the second question, we conducted an empirical investigation into how the algorithm is used in practice. We conducted 13 semi-structured interviews with parents in San Francisco who have used the assignment system and performed content analysis of 12 Reddit threads where parents discussed the algorithm. We complement our qualitative findings with quantitative analysis of application and enrollment data from 4,594 incoming kindergartners in 2017. This triangulation of methods enables us to paint a richer picture of the whole ecosystem in which the algorithm is embedded.

We found that the algorithm failed to support its intended values in practice because it's theoretical promise depended on modeling assumptions that oversimplify and idealize how families will behave and what they seek to achieve. These assumptions overlook the complex barriers to ideal participation that many families face, particularly because of socioeconomic inequalities. Additionally, the system designers vastly underestimated the cost of information acquisition and overestimated the explainability and predictability of the algorithm. In contrast to expectations that the algorithm would ensure an transparent, equitable student assignment process, we find widespread strategic behavior, a lack of trust, and high levels of stress and frustration among families.

Student assignment algorithms promise a clear, mathematically elegant solution to what is in reality a messy, socio-political problem. Our findings show that this clash can not only prevent the algorithm from supporting stakeholders' values, but can even cause it to work against them. Human-centered approaches may help algorithm designers build systems that are better aligned with stakeholders' values in practice. However, algorithmic systems will never be perfect nor sufficient to address complex social and political challenges. For this reason, we must also design systems that are adaptable to complex, evolving community needs and seek alternatives where appropriate.
\section{Related work}

In this work we build on two major areas of related work: work in economics on designing and evaluating matching algorithms for student assignment; and literature in HCI on Value Sensitive Design. We end with a review of literature that examines the role of modeling assumptions in algorithmic systems. In this paper we use the term ``algorithmic system'' or ``student assignment system'' to broadly refer to the matching algorithm as well as the district's processes and families' practices that make up a part of the application and enrollment process. 

\subsection{Matching Algorithms for Student Assignment}

Economists have developed matching algorithms to find optimal assignments between two sides of a market based on each side's preferences \cite{GaleShapley1962, ShapleyScarf1974}. These algorithms have since been applied to numerous real world markets, such as university admissions and organ donation \cite{Roth2015}. Abdulkadiroğlu and Sönmez proposed two variants of matching algorithms\footnote{Deferred Acceptance (DA) \cite{GaleShapley1962} and Top-Trading Cycles (TTC) \cite{ShapleyScarf1974} are both used for student assignment. Student-optimal DA finds the stable matching that most efficiently satisfies student preferences, while TTC finds a matching that is Pareto-efficient in the satisfaction of student preferences but is not guaranteed to be stable.} for assigning students to public schools \cite{Abdulkadiroglu2003}. In these systems, each student submits a ranked list of schools that they would like to attend. Schools may have priority categories for students, such as siblings or neighborhood priorities. Students' preferences are used in conjunction with school priorities to assign each student to an available school seat. These algorithms have promising theoretical properties that should ensure a fair and efficient allocation of seats. For example, they are strategy-proof, meaning students cannot misrepresent their preferences to guarantee an improved outcome. They also produce assignments that efficiently satisfy students' preferences. Student assignment systems based on matching algorithms have been championed for their potential to advance equitable access to high quality education, create more diverse classrooms, and provide more flexibility to families compared to a traditional neighborhood system \cite{Kasman2019}. 

As these systems have been implemented in the real world they have faced new types of challenges, such as confusion for families and decreasing classroom diversity. Pathak suggests that early theoretical literature overlooked or oversimplifed challenges of practical importance \cite{Pathak2017}. Economists have employed empirical methods to further understand strategic behavior \cite{Hassidim2016, Kapor2020, ReesJones2018, Ding2017, Ding2019, Guillen2014, Pais2008, Guillen2017}, information needs \cite{Chen2017, Hermstruwer2019, Guillen2018}, and diversity constraints \cite{Laverde2020, Gonczarowski2019, Hafalir2013, Nguyen2019, Hastings2009, Glazerman2017, Oosterbeek2019}. While these approaches improve some technical shortcomings of the systems, they do not study the values supported by the design of system itself as well as the human factors that shape how it is used in practice.

In this paper we take a human-centered approach and study parents and policy-makers to gain a deeper understanding of their values, attitudes, understandings, and uses of the student assignment system in practice. Kasman and Valant warn that student assignment algorithms are subject to strong political forces and are easily misunderstood \cite{Kasman2019}. They argue that the ultimate success of matching algorithms for student assignment will depend on how people interact with them \cite{Kasman2019}.  Prior work in HCI has studied human values with respect to matching algorithms in experimental settings \cite{Lee2017a, Lee2019}. Central concerns for participants included the algorithms' inability to account for social context, the difficulty of quantifying their preferences, and the lack of opportunities for compromise \cite{Lee2017a}. We build on this work and study stakeholders' values with respect to a high-stakes matching algorithm that has been in use for almost a decade to assign students to public schools. Further, we focus on why the values that these algorithms theoretically support, like transparency and equity, have not been promoted in practice.

\subsection{Value Sensitive Design}

Value Sensitive Design (VSD) is a theoretically grounded methodology to identify and account for stakeholders' values in the design of new technologies \cite{Friedman2006}. In Value Sensitive Design, ``values'' are broadly defined as ``what a person or group of people consider important in life,'' although values with ethical import are considered especially important \cite{Friedman2003}. VSD is a tripartite methodology, involving conceptual, empirical and technical investigations in an iterative and integrative procedure \cite{Friedman2006}. In the conceptual stage, designers identify stakeholders' relevant values. Empirical investigations examine stakeholders' interactions with the technology and how they apprehend values in practice \cite{Davis2015, Friedman2017}. Technical investigations explore how the properties and mechanisms of a particular technology support or hinder values. VSD takes a proactive stance: values should ideally be considered early on and throughout the design process \cite{Davis2015}. However, VSD can also be applied retrospectively to evaluate deployed systems with respect to human values \cite{Friedman2006}. We apply VSD methodology to understand what values San Francisco Unified School District's assignment algorithm was designed to support, and why it has not supported those values in practice, leading to its redesign. 

Zhu et al. adapt the VSD framework to the design and analysis of algorithmic systems through ``Value-Sensitive Algorithm Design'' (VSAD) \cite{Zhu2018}. VSAD emphasizes the need to evaluate algorithms based on whether they are acceptable to stakeholders' values, whether they effectively address the problem they were designed for, and whether they have had positive broader impacts \cite{Zhu2018}. This is in contrast to traditional evaluation procedures for algorithmic systems, which depend heavily on narrow, quantitative success metrics \cite{Zhu2018}. Subsequent work has applied the VSAD framework to reveal stakeholder values in the context of a machine learning algorithm used to predict the quality of editor contributions on Wikipedia \cite{Smith2020}. The authors emphasize the need to integrate values not only into the design of the algorithm itself, but also into the user interface and work practices that form a part of the algorithmic ecosystem \cite{Smith2020}. This is consistent with the interactional principle in VSD, which dictates that ``values are not embedded within a technology; rather, they are implicated through engagement'' \cite{Davis2015}. 

As VSD has been developed and more widely adopted, researchers have encountered some challenges \cite{Davis2015, Borning2012, LeDantec2009}. One challenge is resolving value conflicts, both between stakeholders with different beliefs \cite{Flanagan2005} and between competing values \cite{Shilton2013}. However, even when stakeholders agree on important values, it can be difficult to predict whether a technology that supports a value in theory will actually uphold that value when the system is deployed in the real world. Zhu et al. apply VSAD to design and evaluate an algorithm to recruit new editors to Wikipedia communities \cite{Zhu2018}. They found that their algorithm was acceptable and helpful to the community, but also discovered unanticipated shortcomings. For instance, only more experienced newcomers increased their contributions in response to the recruitment outreach \cite{Zhu2018}. Ames offers another example of values breakdown, contrasting the intended values of the One Laptop Per Child project, such as productivity, with the consumptive values that were enacted in practice \cite{Ames2016}. 

Researchers have identified various causes of  breakdowns between intended values and values in practice. Ames's work highlights the importance of understanding local needs in the context where a technology is to be deployed. Manders-Huits argues that problems can arise when designers misinterpret stakeholders' values, or because stakeholders' values changed over time \cite{Mandershuits2011}. Similarly to this work, Voida et al. find that tension arises from a misalignment between how a computational system operationalizes a value and how the people who use the system understand that value \cite{Voida2014}. We build on these findings by examining a clash between algorithmic logics and real-world goals and practices. We connect these challenges to emerging work studying the role of modeling assumptions and abstraction in algorithmic breakdown.

\subsection{Modeling Assumptions in Algorithmic Systems}

All algorithmic systems rely on an implicit model of the world in order to compute on it. Any model is a simplified abstraction of reality but the simplifying assumptions often go unstated \cite{Box1979}. For example, Selbst et al. describe the \textit{algorithmic frame} in supervised machine learning, in which each observation in labelled training data represents an abstraction of some real-world entity, often a human being \cite{Selbst2019}. The authors warn that algorithmic systems can break down if they rely on abstractions that do not capture important aspects of the interactions between technical and social systems. Researchers have documented challenges both when assumptions are too broad, and when they are overly narrow. For instance, Chancellor et al. identified significant inconsistency in how researchers conceptualize and model humans when using machine learning to predict mental health \cite{Chancellor2019}. In contrast, Saxena et al. found an overly narrow focus on risk prediction in the U.S. child welfare system that oversimplifies the complexity of the domain's needs \cite{Saxena2020}.

In the student assignment context, Hitzig identified how matching algorithms rely on an abstraction of the world that makes strong, unstated normative assumptions regarding distributive justice \cite{Hitzig2019}, or the appropriate distribution of benefits and burdens in a group. The matching paradigm assumes that the ideal outcome is the one where every student is assigned to their first choice school. Hitzig points out that this emphasis on efficiency may not align with school districts' goals, but is often framed in economics as objectively optimal rather than only one of many ways to distribute resources. 

This work demonstrates how unstated, erroneous modeling assumptions about the world can break an algorithmic system. Baumer argues that this breakdown can occur when an algorithm's designers and stakeholders do not share a common understanding of the system's goals and limitations \cite{Baumer2017}. We expand on this work by exploring how the designers of matching algorithms for student assignment relied on certain modeling assumptions about the world in order to justify their designs with respect to values like equity and transparency. We analyze the breakdown of the student assignment algorithm in San Francisco as a case study of what happens when these assumptions clash with stakeholders' real world goals and constraints.

\section{Methods}
Our goal in this research is to understand the values that San Francisco Unified School District's (SFUSD) student assignment system was designed to support and compare and contrast these to parents' experiences in practice. Following Value Sensitive Design methodology \cite{Friedman2003}, we begin with a conceptual investigation drawing on prior literature in economics and SFUSD policy documents to identify the values the system was intended to promote. Then, we conduct a mixed-method empirical investigation to understand why the system ultimately did not support those values and needed to be redesigned. 

\subsection{Data Collection}

We collected data from three sources to understand the district's policy goals (how the system was \textit{intended} to work) and parent experiences (how the system has \textit{actually} worked).

\subsubsection{District Policies}

We collected two official documents from SFUSD to understand the district’s policy goals, their justification for their original design in 2011, and the reasons they voted for a redesign in 2018. We accessed the official policy describing the existing assignment system \cite{policy5101} and the resolution that approved the ongoing redesign \cite{sfusd2018} from the enrollment section of SFUSD’s website.\footnote{\url{https://www.sfusd.edu/schools/enroll/ad-hoc-committee}. Accessed April, 2020.}

\subsubsection{Parent Perspectives}

We collected parent experiences in two primary formats: through interviews with parents, and from public online discussions on social media. The interviews allowed us to ask questions and prompt parents to reflect on and dig deeper into their experiences with the assignment system. The online discussions provide potentially less filtered reflections shared without the presence of researchers and reveal how parents seek and share information online. We supplement this data with a presentation titled ``Reflections on Student Assignment'' by the African American Parents Advisory Council (AAPAC) \cite{aapac}, which was also downloaded from the enrollment section of SFUSD's website.

We conducted semi-structured interviews with 13 parents who have used the student assignment system to apply for elementary schools in SFUSD. We recruited parents through four parenting email and Facebook groups by contacting group administrators who shared a brief recruitment survey on our behalf. During the interview, we asked participants to describe their application and enrollment experiences, and to reflect on their understanding of the assignment algorithm. Interviews were 45 minutes and participants received a \$30 gift card. All interviews were conducted over the phone in English between February and August 2020. 

12 parents completed a demographic survey. Parents reported their income as low income (1), middle income (5), and upper-middle to high income (4) and identified their race or ethnicity as white (4), Asian (3), Chinese (2), white and Hispanic (1), white and Middle Eastern (1), and Vietnamese (1). The 12 respondents reside in six different zip codes in the city. In all 12 households one or more parents had a Bachelor's degree and in nine households the highest level of education was a graduate degree. To preserve participant privacy, we identify participants in this paper by unique identifiers P1 through P13. 

We supplement the interview data with twelve Reddit threads posted on the r/sanfrancisco subreddit\footnote{\url{https://reddit.com/r/sanfrancisco}} between 2016 and 2020. These threads were selected by conducting a comprehensive search of r/sanfrancisco using the search term ``school lottery,'' as it is commonly known to parents.\footnote{Search conducted using the PushShift Reddit repository at https://redditsearch.io/} Each post was reviewed to ensure that it was a discussion of the current SFUSD assignment algorithm. From the twelve threads made up of 678 posts and comments, we manually coded content where the author demonstrated first-hand experience with the assignment algorithm, resulting in a final dataset of 128 posts from 83 contributors. Excluded posts were those that were off topic or presented the author's political view rather than their personal experiences with the system. We paraphrase this content to protect the users' privacy.

\subsubsection{Application and Enrollment Data}
We complement our qualitative data about parent experiences with publicly available, de-identified kindergarten application data from 2017 to understand higher-level trends in how parents use the system.\footnote{The data was collected as part of a public records request by local journalist Pickoff-White for a story about how parents try to game the system \cite{White2018}. The data is available at \url{https://github.com/pickoffwhite/San-Francisco-Kindergarten-Lottery}.} For each of the 4,594 applicants, the data includes their ranked list of schools, the school they were assigned to, and the school they enrolled in. It also includes the student's zipcode, race, and whether the student resides in a census tract with the lowest performing schools (CTIP1 area), which makes them eligible for priority at their preferred schools. Applicants are 28\% Asian or Pacific Islander, 24\% white, 23\% Hispanic and 3.2\% Black. 21\% declined to state their race. Approximately 15\% of applicants were eligible for CTIP1 priority, 45\% of whom are Hispanic. 11\% of CTIP1-eligible students are Black, which is 53\% of all Black applicants.

\subsubsection{Limitations}
We recruited interview participants through convenience sampling online and complemented the interviews with existing online data, which biases our data towards those who have the time and motivation to participate in research studies, online discussions, and district focus groups. Our dataset lacks sufficient representation of low-income families and Black and Hispanic families. It is important that future work addresses this limitation, particularly considering that integration is a key goal for the school district, and that these families are underrepresented in existing discourses. In future work we will focus on understanding the experiences of historically underserved families with student assignment algorithms, specifically families of color, low-income families, and families with low English proficiency. 

\subsection{Data Analysis}

In order to understand the district's values for student assignment and the reasons why the assignment algorithm has not supported these values, we conduct inductive, qualitative content analysis \cite{qualresearch} and quantitative data analysis.

\subsubsection{Qualitative Analysis}

Our qualitative dataset was made up of district policy documents and community input, interview transcripts, and Reddit content. We performed an open-ended inductive analysis, drawing on elements of grounded theory method \cite{groundedtheory}. We began with two separate analyses: one to understand the district’s values and policies; and a second to understand parent experiences and perspectives. The authors met regularly throughout the analysis to discuss codes and emerging themes. In both analyses we began by conducting open coding on a line-by-line basis using separate code books \cite{groundedtheory}. We then conducted axial coding to identify relationships between codes and higher level themes. In the axial coding stage for the SFUSD policy documents, we identified three high level codes relevant to our research questions: Values: What are the district's values and goals for student assignment?; Mechanism: How was the district's current system expected to support their values?; and Challenges: Why did the district ultimately decide to redesign the system?. Next, we analyzed parent perspectives from the community input documents, interview transcripts, and Reddit content. We conducted two rounds of open coding. First, we focused only on these three data sources. We identified codes that included "priorities," "algorithmic theories," and "challenges." Then, we linked the open codes from the first round to the challenges identified in the policy documents. We found that challenges parents described in our parent perspectives dataset were relatively consistent with those described in the policy documents and we reached theoretical saturation after approximately ten interviews. 

\subsubsection{Quantitative Analysis}
We linked the application dataset to publicly available school-level standardized test results in order to understand how families use the system to access educational opportunities. We accessed third grade results in the California Smarter Balanced Summative Assessments in 2017-2018, provided by the California Department of Education.\footnote{Data available at url{https://caaspp-elpac.cde.ca.gov/caaspp/ResearchFileList}. We link the school achievement data to the applications by state-level (CDS) code. The preference data contains only school numbers, a district-level coding scheme. SFUSD has published a document linking these district school numbers to the school name and state-level (CDS) codes \url{http://web.sfusd.edu/Services/research_public/rpadc_lib/SFUSD\%20CDS\%20Codes\%20SchYr2012-13_(08-20-12).pdf}.} We conducted exploratory data visualization to investigate trends in preferences. We measure variation in preferences by race and CTIP1 priority status in order to gain insight into if and how participation varies across groups differently impacted by structural oppression and historical exclusion from high quality education. We present quantitative findings using visualizations to include all students. When comparing summary statistics we use the bootstrap\footnote{We use percentile intervals to estimate confidence intervals and the bootstrapped t-test to estimate p-values for differences in means using 10,000 re-samples, following \cite{bootstrap}. Groups (race and CTIP1) are re-sampled independently.} method to estimate statistical significance \cite{bootstrap}. For this analysis we used third grade standardized test results as a rough estimate of resources and opportunities at each elementary school. We recognize that there are many ways in which schools provide value to children that are not reflected in standardized test results.

\section{Student Assignment in San Francisco: Intended Values}

In this section, we present our findings on the values that San Francisco Unified School District (SFUSD) intended their student assignment system to support. In the next section we analyze why this system did not realize those values in practice. 

SFUSD has been utilizing different choice-based systems to address educational inequality in the district for almost forty years \cite{sfusd2015}. Although the mechanism for assigning students to schools has changed significantly over time, SFUSD has been consistent in their values and goals for student assignment. Their current policy designates three primary goals:

\begin{enumerate}
    \item ``Reverse the trend of racial isolation and the concentration of underserved students in the same school;
    \item Provide equitable access to the range of opportunities offered to students; and
    \item Provide transparency at every stage of the process.'' \cite{policy5101}
\end{enumerate}

In addition, they emphasize the importance of efficiently utilizing limited district resources, ensuring predictability and ease of use for families, and creating robust enrollments at all schools.

\begin{figure*}
    \centering
    \includegraphics[width=\textwidth]{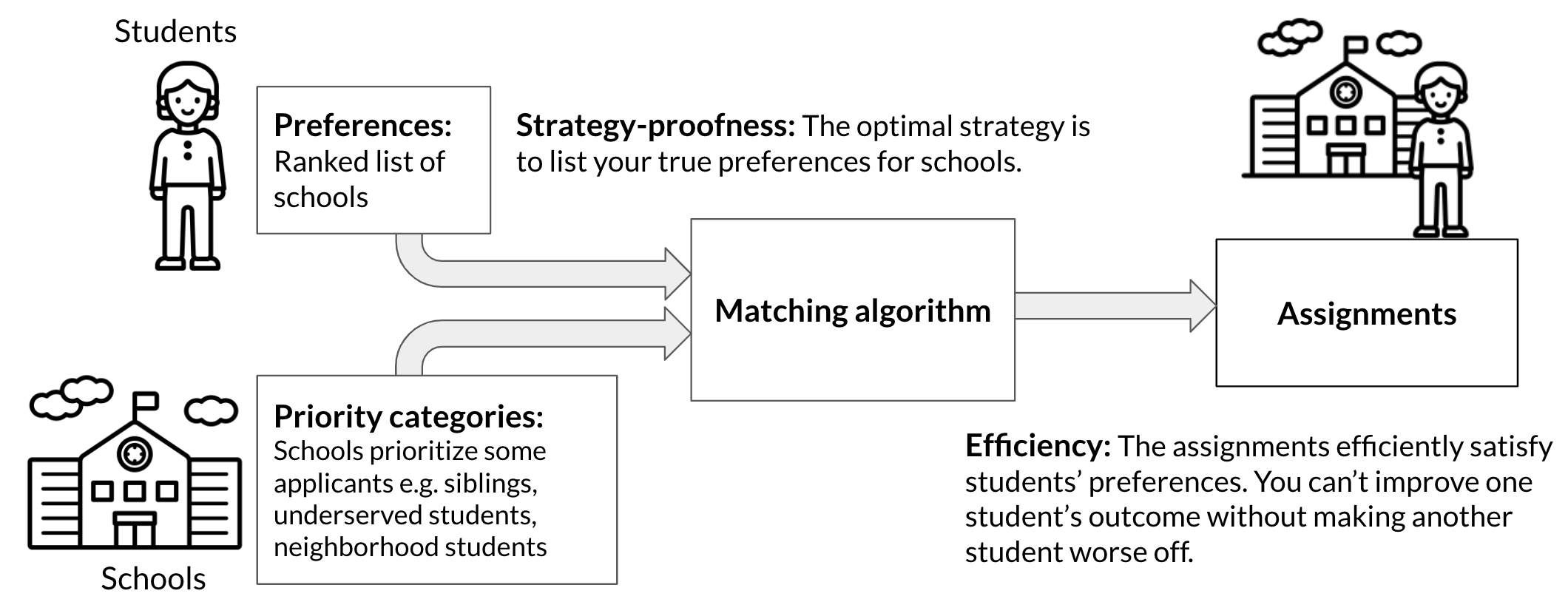}
    \caption{The matching algorithm takes students' preferences over schools and schools' pre-defined priority categories as inputs and outputs the most efficient assignment of students to schools.}
    \Description[A flow chart showing the inputs (student preferences and school priorities) to the matching algorithm and its output (student-school assignments), labelled with key properties: strategy-proofness and efficiency.]{A flow chart showing the inputs and outputs to the matching algorithm with key properties labelled. The flow chat is identical to Figure \ref{fig:splash}. The first label to the right of the preferences box reads, ``Strategy-proofness: The optimal strategy is to list your true preferences for schools.'' The second label is to the right of the matching algorithm box and it reads, ``Efficiency: The assignments efficiently satisfy students' preferences. You can't improve one student's outcome without making another student worse off.''}
    \label{fig:model}
\end{figure*}

In SFUSD's current assignment system \cite{sfusd2015}, students or their parents apply for schools by submitting their \textit{preferences}: a ranked list of schools they would like to attend (Figure \ref{fig:model}). To increase flexibility and access to opportunities, students can rank any school in the district and there is no limit on the number of schools they can rank. The district also defines priority categories. Elementary schools give top priority to siblings of continuing students and then to underserved students. Underserved students are defined as those living in neighborhoods with the schools that have the lowest performance on standardized tests, known as \textit{CTIP1} areas. The matching algorithm\footnote{SFUSD uses a variant of the Top Trading Cycles algorithm \cite{ShapleyScarf1974}. See \cite{Abdulkadiroglu2003} for a technical analysis of Top Trading Cycles in the student assignment context or \cite{Roth2015} for a more broadly accessible introduction to market design.} then takes student preferences and school priorities and produces the best possible assignments for the students subject to the schools' priorities and capacity constraints. Importantly, the resulting assignments from this algorithm are guaranteed to efficiently satisfy \textit{student preferences} not school priorities. School priorities are only used to determine which students are assigned to over-demanded seats. The matching algorithm is also strategy-proof, meaning that it can be theoretically proven that families do not benefit from manipulating their preferences to game the system.

We consolidated the school district's stated goals for student assignment into four high-level values: (1) transparency, predictability and simplicity; (2) equity and diversity; (3) quality schools; and (4) community and continuity (Table \ref{overviewtbl}). In this section, we described the system that was expected to support these values. In the next section, we explore why these expectations were not met in practice.

\section{Algorithmic Breakdown: Values in Practice}

In December 2018, San Francisco Board of Education determined that the the algorithm was not working as intended \cite{sfusd2018}. While the number one stated goal of the algorithm was to ``reverse the trend of racial isolation and the concentration of underserved students in the same school,'' the Board found that segregation had \textit{increased} since the algorithm was introduced and there was widespread dissatisfaction amongst parents \cite{sfusd2019, policy5101}. The assignment algorithm had failed to respect the values that it was designed to support and the Board voted to stop using it and to design a new system. In this section we present our findings that help explain why. 

For each of the district's four high-level values for student assignment (Table \ref{overviewtbl}), we first review the theoretical properties and promises of the algorithm related to that value: why would economists and district policy-makers expect that the system would respect that value? Next, we analyze what implicit modeling assumptions those expectations depend on. Finally, we explain how families' needs, constraints, and values in the real world clashed with system designers' assumptions about them, which prevented the algorithm from meeting its theoretical promises and enacting the district's values in practice.\footnote{In this work we identify the school district's values and draw on families' experiences to explain why they haven't been supported. The district's values may not completely align with families' values. We assume that satisfying families is one of the district's priorities, and we find substantial overlap between the four district values and what parents in our sample find important. We leave a detailed analysis of families' values to future work.} 

\begin{table*}
\caption{We consolidated the San Francisco Unified School District's goals for student assignment into four overarching values. Assignment algorithms have theoretical properties aligned with these values. However, the San Francisco assignment algorithm's theoretical promises have not been realized because they rely on modeling assumptions that clash with real world challenges.}

\label{overviewtbl}
\begin{tabular}{p{0.15\textwidth} p{0.2\textwidth} p{0.275\textwidth} p{0.275\textwidth}}
    \toprule
     \textit{Value} & \textit{Promises and Properties} & \textit{Modeling Assumptions} & \textit{Real World Challenges}\\
    \midrule
    Transparency, & Algorithm has a clearly & The district provides accessible, & Finding and understanding \\
    predictability, and & defined procedure. & clear information. Families want and & information is difficult. Some parents \\
    simplicity &  Assignments are explainable. & understand explanations. Families do not try to game the system. & try to game the system. There is a lack of trust: assignments are perceived as unpredictable and unfair. \\
    \midrule
    Equity and diversity &  
        Any student can apply to any school. Underserved students are given priority access. &
        All families participate equally and the all-choice system offers identical opportunities to all families. &
        Time, language and economic constraints create participation barriers for lower resourced families. \\
    \midrule
    Quality schools & 
        Competition for applicants will drive up the overall quality of schools in the district. &
        Families base their preferences on accurate estimates of school quality. Schools can respond to competitive pressures. &
        Competition is driven by social signalling and negative stereotypes. Underserved schools lack resources to attract applicants.\\
    \midrule
    Community and & Priority for siblings & Schools have sufficient capacity to & A lack of guaranteed access to \\
    continuity & 
        and students in the school's attendance area. &
         handle demand from siblings and neighborhood children. &
        local schools frustrates families living in neighborhoods with very popular schools.\\
    \bottomrule
\end{tabular}
\Description[Table summarizing the main findings of the paper.]{Table summarizing the main findings of the paper. There is a header and followed by four rows of data, one corresponding to each of the four high-level school district values for student assignment. The first column contains the value. The second column summarizes the promises and properties of the assignment algorithm that were supposed to support the value. The third column summarizes the modeling assumptions that these promises and properties depend on. The fourth column summarizes the real world challenges that clash with the modeling assumptions.}
\end{table*}

\subsection{Transparency, Predictability, and Simplicity}

\subsubsection{Theoretical promises}

Matching algorithms are clearly and explicitly defined procedures. This differentiates them from assignment systems based on imprecise admissions criteria, which have historically been more difficult to justify and have led to legal disputes \cite{Abdulkadiroglu2003}. If a student wants to understand why they did not receive an assignment they were hoping for, the algorithm’s decision can be explained. Matching algorithms are also provably strategy-proof. That is, students cannot guarantee a more preferable assignment by strategically misrepresenting their preferences. Strategic behavior requires time and effort, so preventing strategic advantages is critical not only for simplicity and efficiency, but also for ensuring that all families can participate equally.

\subsubsection{Modeling assumptions: families will accept their assignment as fair and legitimate as long as the algorithm's logic is explained to them.}

This assumes that the school district provides an accessible, comprehensible explanation and that families would seek out, understand, and trust this explanation. Families have known preferences for schools and recognize that they should report those preferences truthfully.

\subsubsection{Real world challenges}

In practice, families find the assignment system difficult to navigate and struggle to find relevant, clear, and consistent information. Some parents engage in strategic behavior to try to improve their child's assignment, contrary to theoretical incentives. Rather than seeking and accepting an explanation, families who are dissatisfied with their assignment seek to change it. Families' trust in the system is eroded by the lack of clear information and the belief that some parents are able to game the system.

Parents face a significant information cost to understand the various opportunities available across the city. There are 72 elementary school programs in SFUSD \cite{sfusd2018}. Parents indicated that researching schools is a burdensome time-commitment. In-person school visits are a popular source of information when forming preferences, but these visits are time-intensive and logistically difficult. 

\begin{quote}
    \textit{[\ldots I]t's like a full time job doing all the school tours. (P7)}
\end{quote}

Online information is another widely used source, but school information is not centralized, nor is it consistent across schools. A number of parents mentioned the difficulty of navigating online district resources:

\begin{quote}
    \textit{[\ldots F]inding and gathering the information about the schools from the district is a mess. (P11)}
\end{quote}

None of the parents we interviewed felt that they had a clear understanding of how the algorithm works. The algorithm is colloquially known to parents as ``the lottery.'' Although the algorithm has only a small lottery aspect to break ties between students with the same priority, many believe it is mostly or entirely random.

\begin{quote}
    \textit{I’m not really that confident in their actual lottery system. It could be bingo in the background for all I know. (P4)}
\end{quote}

This leaves families feeling a lack of agency and control over their child's education.

\begin{quote}
    \textit{I mean, the word itself, lottery, most of it is random. I don’t feel like we can do anything at all. (P5)}
\end{quote}

Confused and frustrated by district resources, parents frequently seek advice from other parents online and in-person. Reddit users sought and shared complex strategies, sometimes relying on substantial independent research. This is consistent with prior work showing that advice sharing in social networks can encourage strategic behavior \cite{Ding2017, Ding2019}. Advice from other families is often conflicting and unclear, further exacerbating confusion about the system.

\begin{quote}
    \textit{[W]e also got different advice from different parents. They're very, very different from each other. Some people say, ``Put in as many schools as possible,'' and some people say, ``No, just put two schools that you really wanted, and then you have a higher chance of getting those.'' (P5)}
\end{quote}

\begin{figure*}
    \centering
    \includegraphics[width=\textwidth]{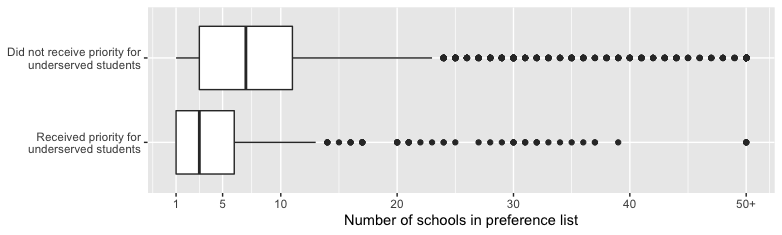}
    \caption{Families who were eligible for priority for underserved students ranked fewer schools on average (mean: 5.5; 95\% CI: 5.0-6.2) than other families in the city (mean: 11.6; 95\% CI: 11.2-12.1; difference in means: p=0.00). This may suggest that stategic behavior is more widespread amongst higher resource families.}
    \Description[A boxplot showing the distribution of application length by underserved student priority status. Underserved students submitted shorter applications, overall.]{A boxplot showing the distribution of application length for students who received priority for underserved students and those who did not. The median for underserved students is 3 schools and the interquartile range is 5. For other students the median is 7 schools and the interquartile range is 8. Both boxplots are skewed to the right with outliers with very long applications. This tail is longer and heavier for advantaged students.}
    \label{fig:listlen}
\end{figure*}

The 2017 application data indicates that strategic behavior may be more widespread amongst more privileged families. On average, families who were eligible for the CTIP1 priority for underserved students ranked 5.5 schools in their application (95\% confidence interval (CI): 5.0--6.2 schools), while families in other areas of the city ranked an average of 11.6 (95\% CI: 11.2--12.1 schools; difference in means: p = 0.00)  (Figure \ref{fig:listlen}). 96\% of families eligible for CTIP1 priority were assigned their first choice, so this difference may reflect these families' confidence that they will get one of their top choices. On the other hand, it may reflect disparities in access to the time and resources needed to research schools and strategies. White students submitted especially long preference lists (mean = 16.5; 95\% CI: 15.6--17.6),\footnote{Differences in means between white students and Black, Asian or Pacific Islander, and Hispanic students is highly statistically significant, even with conservative adjustments for multiple hypothesis testing.} a further indication that strategic behavior is more popular with families with more structural advantages.

Receiving an unfavorable assignment was a major concern for families in our sample. The district offers multiple rounds of the assignment algorithm, which many  parents participate in if they are dissatisfied with their child's assignment. However, this process can be long, uncertain, and frustrating. Some parents received the same assignment every round with no further explanation or assistance.

\begin{quote}
    \textit{[\ldots T]he first announcement that we got [\ldots], I actually wasn't that upset. I said, ``You know what, there's more rounds. [\ldots] We could stick it out.'' But I was really upset at the second one because there was literally no change. And that really had me questioning, ``I'm just trying to play by the rules. Should I not trust this any more than it's going to work out?'' (P9)}
\end{quote}

Parents on Reddit recommended unofficial avenues for recourse, many of which require substantial time and resources. These include going in person to the enrollment office repeatedly to request reassignment, remaining on waiting lists up to ten days into the school year, and opting out of the public school system altogether. 

Overall, a complicated algorithm together with a shortage of transparent and accessible information has fostered distrust and frustration amongst parents in the district. Distrust is fuelled by perceptions that the system is random and unscientific, and that it allows parents with more time and resources to gain an unfair advantage.

\begin{quote}
    \textit{It's definitely convoluted. It's definitely multilayered, it's complex. And that favors people who have the time and the wherewithal to figure it out. [\ldots T]he complexity invites accusations of [corruption] and does not inspire trust (P9)}
\end{quote}

\subsection{Diversity and Equity}

\subsubsection{Theoretical promises}

The assignment system is an all-choice system with unrestricted preference lists, so any student can apply to any school in the district. Compared to a neighborhood system, or even more restricted choice systems, this design has the potential to enable more equitable access to educational opportunity. In an effort to promote equitable access to education and diverse schools, SFUSD has added the CTIP1 priority category, which gives priority admission at over-demanded schools to students from neighborhoods with under-performing schools.

\subsubsection{Modeling assumptions: all families participate equally in the system and the all-choice system offers identical opportunities to all families.} CTIP1 students prefer to attend over-demanded schools if they can access them. Applicant pools reflect the racial and socioeconomic diversity of the city.

\subsubsection{Real world challenges}

Although an all-choice system offers greater flexibility than a neighborhood system, our results show that families with fewer resources face significant barriers to ideal participation in SFUSD’s choice system. Although families can rank any school on their application, some families are not able to choose the schools that offer the greatest opportunities. Preferences are segregated by race and income, preventing the algorithm from creating diverse assignments.

Our results indicate that the all-choice system does not offer identical opportunities to all families. Every family can apply to any school, but that does not mean that every family can actually access every school. For example, transportation logistics can be a significant challenge. When choosing a kindergarten for their child, P1 met with an education placement counselor at SFUSD to understand the special education services offered across the district. P1 recalled their response to one of the counselor’s suggestions:

\begin{quote}
    \textit{So, you are telling me this school is [\ldots] three blocks uphill and we're supposed to do that with a kindergartner and no car? [\ldots] There's no way that on my worst day that I would be able to drag my kindergartner with special needs uphill in the rain. (P1)}
\end{quote}

\begin{figure*}
    \centering
    \includegraphics[width=\textwidth]{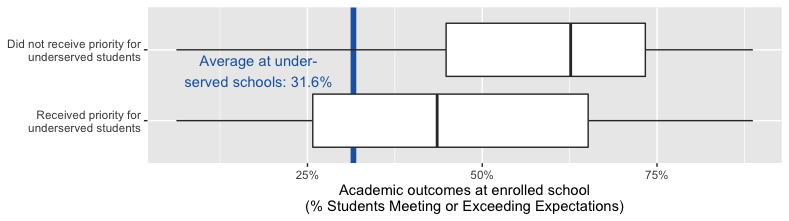}
    \caption{The priority for underserved students helps those students access educational opportunity, but there remain inequities that priority enrollment cannot address. Students with priority enrolled in higher performing schools (mean: 45.0\% of students met or exceeded expectations on standardized tests; 95\% CI: 43.2\% -- 46.7\%), than their average neighborhood school (mean: 31.6\%). However, they still enrolled in lower performing schools on average than students who were not eligible for priority (mean: 57.2\%; 95\% CI: 56.5\%--57.9\%) (difference in means: p = 0.00). Academic outcomes are measured as the percentage of third grade students at the enrolled school who met or exceeded expectations in the 2017-18 statewide assessments.}
    \Description[A boxplot showing that underserved students enroll in lower performing schools than other students.]{A boxplot showing the distribution of standardized test performance at schools where underserved students enrolled compared to other students. The median for underserved students is 43.55\% and the interquartile range is 39.4\%. For other students the median is 62.65 \% and the interquartile range is 28.5\%. The average across underserved students' neighborhood schools is shown with a vertical reference line at 31.6\% which is labelled ``Average at underserved schools.''}
    \label{fig:outcomes}
\end{figure*}

The CTIP1 priority is potentially a useful advantage for underserved students. In 2017, 96\% of students who were eligible for this priority were assigned their first choice school, compared to 58\% of students without this priority. However, CTIP1 priority is only useful for advancing educational equity if these students can actually use it to enroll in well-resourced schools. In 2017, students with CTIP1 priority enrolled in schools with lower academic outcomes than other students (Figure \ref{fig:outcomes}). On average, underserved students enrolled in a school where 45.0\% of third graders met or exceeded expectations in the English Language Arts/Literacy exams\footnote{Qualitatively similar results to those presented in this section hold for Mathematics results.} (95\% CI: 43.2\% -- 46.7\%), compared to 57.2\% (95\% CI: 56.5\% -- 57.9\%) of students at the average school that other students enrolled in (difference in means: p = 0.00). This difference points to persisting inequities in access to higher resource schools that priority assignment is insufficient to address. CTIP1 priority cannot, for example, help students access schools that are physically inaccessible for them. Social factors may also influence choice patterns. For instance, the African American Parent Advisory Council (AAPAC) has raised concerns that Black students in San Francisco continue to face racism and biases in racially diverse classrooms \cite{aapac}. 

These findings are consistent with prior work showing that while proximity and academics are important to most families, more privileged parents tend to put more emphasis on a school’s academic performance \cite{Hastings2009, Abdulkadiroglu2019, Burgess2015}, while parents from low-income or racialized backgrounds may be more likely to prioritize proximity \cite{Laverde2020} or representation of students from a similar background \cite{Hastings2009}. As a result of differences in students' preferences, applicant pools at schools across the city are segregated by race and income. This prevents the algorithm from creating diverse assignments \cite{sfusd2018, Laverde2020}.

\subsection{Quality Schools}

\subsubsection{Theoretical promises}

System designers have suggested that choice systems indirectly improve school quality. For instance, Pathak argues that matching mechanisms create competition between schools, which pushes under-demanded schools to improve in order to attract applicants and sustain their enrollment \cite{Pathak2017}. In addition, Pathak points out that an algorithmic system based on student preferences creates a useful source of demand data for the district to target interventions or closures at underenrolled schools \cite{Pathak2017}.

\subsubsection{Modeling assumptions: a competitive market will drive up the overall quality of offerings.}
This assumes that demand is driven by accurate estimates of school quality.

\subsubsection{Real world challenges}

Unfortunately, competition in SFUSD has not resulted in an improvement in educational opportunities and outcomes across the district \cite{sfusd2018}. Our findings reveal that parents base their preferences on noisy signals of school quality. Still, some students depend on under-demanded schools and are harmed by under-enrollment and school closures.

Our results suggest that parents' preferences are strongly shaped by social learning and stereotypes. Many parents reported using other parents' opinions and experiences of schools to inform their preferences. Some feel that a few schools are disproportionately regarded as the ``best'' schools in the city. Parents on Reddit attested that many good schools are unfairly dismissed by more advantaged parents, sometimes on the basis of thinly veiled racist and classist stereotypes. Standardized test scores or aggregate scores like those reported by greatschools.org are another popular source of information. Though seemingly more objective, these measure are heavily correlated with resources and demographics at schools \cite{barnum2019}, further exacerbating preference segregation. In the presence of these types of competitive pressures, well-resourced schools are heavily over-demanded while under-resourced schools struggle to maintain robust enrollments \cite{sfusd2018}. SFUSD believes the algorithm has created ``unhealthy competition'' between schools, resulting in schools ranging in size from 100 to nearly 700 students \cite{sfusd2019}. 

While Pathak argues that choice patterns are useful in determining which schools to close and which to support and expand \cite{Pathak2017}, this overlooks the correlation between demand patterns and existing patterns of inequality. Under-enrollment and school closures can seriously harm the communities at those schools, which often serve predominantly poor students of color \cite{schoolcolors, ewing2018}. SFUSD has acknowledged the need to more equitably distribute resources, but it can be politically difficult to direct resources to schools with low demand and enrollment \cite{policy5101}.

\subsection{Community and Continuity}

\subsubsection{Theoretical promises}

SFUSD’s sibling and attendance area priority categories are designed to encourage a sense of community and cohesion for families. In addition, students attending PreK or Transitional Kindergarten in the attendance area are given priority to ensure continuity for students.

\subsubsection{Modeling assumptions: schools have sufficient capacity to handle demand from siblings and neighborhood children.}

\subsubsection{Real world challenges}

Many families are dissatisfied by a lack of access to their local schools. In many neighborhoods there is a mismatch between demand for the attendance area school and its capacity. In fact, current attendance area boundaries are drawn such that some schools do not have the capacity to serve every student in the attendance area \cite{sfusd2015}. As a result, the attendance area priority does not provide an acceptable level of predictability for those who want to enroll in their local school.

For parents living in neighborhoods with popular schools, access to their attendance area school is far from guaranteed. One Reddit user expressed frustration after they found out that they may not be able to enroll their child in their local school. Due to their family's circumstances, they feared it would be impossible to get their child to a school further from home.

Parents in our sample value access to local schools for convenience and a sense of community. Under the existing system, two children who live close to each other may attend schools on opposite sides of the city. There are even neighborhoods in San Francisco where students are enrolled across all 72 elementary school programs \cite{sfusd2018}. Some parents felt that this dispersion undermines the educational experience for children:

\begin{quote}
    \textit{[I]t is really important for our children to bond and build relationships in their community. And they really connect to their education and their educational environment very differently [when they do]. (P1)}
\end{quote}

By underestimating the mismatch between demand for neighborhood schools and capacity at those schools, the assignment system has generated significant dissatisfaction among parents who live near popular schools. These parents are increasingly pushing for a return to a traditional neighborhood system. However, this would restrict flexibility and access to educational opportunities for many families across the city who use the system to enroll their children in schools other than their neighborhood school.\footnote{A district analysis showed that 54\% of kindergarten applicants did not list their attendance area school anywhere in their preference list for the 2013-14 school year \cite{sfusd2015}. This is especially true of underserved students: according to the 2017 application data, around 75\% of students who received CTIP1 priority enrolled in an elementary school outside of the CTIP1 census tracts. Schools in CTIP1 census tracts were determined according to the definition updated for the 2014-15 school year \url{https://archive.sfusd.edu/en/assets/sfusd-staff/enroll/files/Revising_CTIP1_for_2014_15_SY.pdf}.}

\section{Design Implications for Student Assignment}
 
In the previous section we showed how incorrect or oversimplified modeling assumptions have played a role in the breakdown of the student assignment algorithm in San Francisco. In this section we draw on these findings to present four design implications for student assignment systems: (1) provide relevant and accessible information; (2) (re)align algorithmic objectives with community goals in mind; (3) reconsider how stakeholders express their needs and constraints; and (4) make appropriate, reliable avenues for recourse available. We emphasize that student assignment is a complex, socio-political problem and our results and recommendations are our first step to better understanding it. In the future, we will continue this work focusing explicitly on the needs of underserved students. In the next section we discuss broader implications of this work for the design of algorithmic systems.

\subsection{Provide relevant and accessible information}
When looking for a school for their child, parents need to find schools that meet their needs, and then understand how to apply. Our research shows that information acquisition is very difficult, which leaves families with a sense of distrust and perceptions of randomness, unpredictability, and unfairness. However, more information is not always better. Information about algorithmic systems should be congruent with stakeholder needs and interests and should be limited to the most relevant information in order to minimize cognitive load \cite{Dietz2003}. In the student assignment setting, we found the most salient information for families is information about the schools available to them that best meet their needs. Relevant, accurate information should be easy to find and navigate. San Francisco Unified School District has recognized this need and has committed to making this information available in a variety of languages \cite{sfusd2018}. Further work is needed to understand what kind of information about schools will be relevant and helpful without exacerbating negative stereotyping and preference segregation.

Transparency information about the algorithm itself may also reduce stress and increase trust in the system, but only if this information is clear and useful \cite{Nissenbaum2011, Kulesza2013, Cheng2019}. The algorithmic information most relevant to parents in our sample is their chances of receiving a particular assignment. This information is currently difficult to find, in part because these probabilities depend on others’ preferences. However, this information may reduce stress and increase predictability. One concrete goal moving forward could be to ensure that information about schools and admission probabilities are easily available.

\subsection{(Re)Align algorithmic objectives with community goals in mind}

SFUSD expected their assignment system to satisfy individual preferences \textit{and} promote community-level goals like equitable access to education and diverse classrooms. However, the system has had limited success in promoting educational equity, and racial and economic segregation has worsened since it was introduced \cite{sfusd2018}. One reason for this breakdown is that the primary objective of matching algorithms is to efficiently satisfy students' preferences, and in San Francisco students' preferences are already heavily segregated by race and income \cite{sfusd2018}. This indicates a breakdown between community goals and what the algorithm is optimizing for.

The focus on satisfying students' preferences can also obscure other problems. For example, if we look only at preference satisfaction, then underserved students appear to have a strong advantage in the current system. 96\% of incoming kindergartners who were eligible for priority for underserved students received their first choice school in 2017, compared to only 58\% of other students. However, underserved students continue to enroll in lower resourced schools and an opportunity gap persists between underserved students and others in the district. Due to the limitations of our sample, we cannot conclusively explain the reasons for segregated and unequal preferences. Nevertheless, these two challenges suggest that technical system designers need to work closely with policy-makers and community members to ensure that their algorithm's objectives and evaluation metrics are aligned with higher-level goals and values. 

\subsection{Reconsider how stakeholders express their needs and constraints}

Another way to make progress towards community goals is to reconsider how families express their values, needs, and constraints. Matching algorithms model families as independent, self-interested agents with some inherent preferences over schools. Schools are assumed to be merely ``objects to be `consumed' by the students'' \cite{Abdulkadiroglu2003}. However, our findings highlight that preferences are based on limited information and are strongly shaped by social context. Schools are also important communities for children and their families. Researchers have found that matching algorithms for group decision-making do not give participants the space to understand each others’ concerns and arrive at compromises that might be natural in a negotiation amongst humans \cite{Lee2017a, Lee2017b}.  One avenue for future work is to develop alternative methods for eliciting students' preferences that better reflect their needs and allow for compromise and community building. For example, families could submit their weighted priorities over factors like language programs or proximity to their home. In our interviews we found that parents already make these types of comparisons frequently when researching schools. Such an approach might help shift families' focus from how high their assigned school was in their personal ranked list to how their assigned school meets their needs and constraints and contributes to progress towards community-level goals. 

\subsection{Make appropriate, reliable avenues for recourse available}

Because there is limited space at popular schools, some students will receive a disappointing assignment. There are multiple rounds of the algorithm for students who wish to appeal their assignment. However, our results suggest that this process can be frustrating and unpredictable. One concrete recommendation is to improve communication with parents throughout the process about their application status and their available next steps. Our findings also suggest that privileged stakeholders will continue to seek unofficial channels to achieve their goals. Therefore, future work developing fair processes for recourse should prioritize the needs of lower resourced stakeholders and design low cost appeals processes.

\section{Discussion and Future Work}

In the previous section, we suggested ways to improve student assignment algorithms to better support stakeholders' values. In this section, we discuss the implications of our work for algorithm design more broadly and identify opportunities for future work. 
 
This work presents an example of how incorrect assumptions can prevent a system from supporting intended values in practice. Direct engagement with stakeholders early on in the design process may help system designers identify incorrect or oversimplified modeling assumptions. For example, economists initially assumed that matching algorithms would be easy to explain to families and that the procedure would be perceived as fair.  A value sensitive approach would have encouraged designers to engage with stakeholders early in the development process to gauge their perceptions and acceptance of the technology \cite{Zhu2018}. Economists may have discovered that stakeholders’ acceptance of matching algorithms for student assignment would depend heavily on social and political factors, such as pre-existing institutional trust in the school district. 

Even with improved methods to align algorithm design with stakeholders’ values, unanticipated challenges will arise because algorithmic systems must rely on \textit{some} abstractions and assumptions that will always be an imperfect approximation of the real world \cite{Box1979, Selbst2019}. Crawford analyzed sites of conflict between algorithms and humans, and has warned of the danger of understanding algorithmic logics as autocratic \cite{Crawford2016}. Instead, algorithmic systems should be accountable to community values beyond the formal design process and stakeholders should have ongoing opportunities to voice concerns, even after the system has been deployed \cite{Zhu2018}.  Future work is needed to design algorithmic systems that are adaptable and flexible in response to this feedback. 
 
In advocating for ongoing engagement with stakeholders, it is important to grapple with differences in power and participation among them \cite{Zhu2018, Dietz2003}. We need to design mechanisms for participation that are equitable and low-cost for lower resource families to voice their concerns \cite{Harrington2019}. In the student assignment setting, we found that convenience sampling strongly skewed our sample of parents towards higher resource parents with the time and motivation to voice their concerns. While building a system that serves all stakeholders is ideal, trade-offs are inevitable when systems impact a large number of stakeholders with diverse perspectives and needs \cite{Zhu2018, Dietz2003}. Avenues for participation should encourage deliberation of trade-offs and include safeguards to prevent powerful stakeholders from compromising important community values in order to design a system that better serves their own interests.

Designing systems while taking into account stakeholders with conflicting values and priorities will require a broader view of algorithmic performance. The research literature on matching algorithms has typically emphasized theoretical guarantees, such as whether assignments are efficient or stable. A human-centered analysis of algorithmic performance would involve evaluating the system in its real world context, along dimensions such as acceptance from stakeholders and broader impacts \cite{Zhu2018}. This is in contrast to typical practices in algorithmic fields such as machine learning, where algorithms are developed and evaluated with respect to narrow, quantitative metrics such as efficiency. A broader view of algorithmic performance may identify challenges that are central to stakeholders' experiences with the system if not directly related to the algorithm's design, such as the difficulty of forming a preference list.

Finally, we cannot expect that every algorithmic system can support community values if only the right design choices are made. Demand for a technology in the first place is often closely tied to particular politics, which may necessitate certain values and preclude others. For example, education researcher, Scott argues that modern school choice programs reflect a neoliberal ideology focused on empowering parents as consumers of educational opportunities for their child \cite{Scott2013}. Advocates claim that school choice promotes educational equity by enabling underserved students to attend a school other than their neighborhood school. Assignment algorithms can support this approach to equity with technical features like priority categories or quota systems. However, this is not the only approach to educational equity. In fact, it offers limited benefits to those who do not have the time or resources to exercise informed choice \cite{Scott2011a}. A redistributive principle, on the other hand, would prioritize providing underserved students with educational opportunities in their own communities and protecting local students' access to those resources \cite{aapac}. Assignment algorithms cannot effectively support such an approach: increasing enrollment at under-demanded schools using an algorithm would require violating some students' preferences and may be disruptive and harmful to the existing communities at those schools \cite{aapac, schoolcolors2}. Therefore, student assignment algorithms exist within and to uphold a political ideology that privileges individual choice sometimes at the cost of other values, such as democracy, resource equality, and desegregation \cite{Scott2011a}. This example shows why it is important not only to consider how certain design choices \textit{might} support the values that stakeholders find salient, but also what values a technology \textit{necessitates or precludes} based on the implicit politics of its existence. Value Sensitive Design does not provide an explicit ethical theory to designate what kinds of values \textit{should} be supported \cite{Mandershuits2011, Borning2012}. Therefore, in addition to an understanding of implicit values and politics, our analysis must include a commitment to justice \cite{costanza2018design} and accept refusal as a legitimate way of engaging with technology \cite{cifor2019feminist}.

\section{Conclusion}

In this paper we conduct qualitative content analysis of parent experiences and district policies, and quantitative analysis of elementary school applications to understand why the student assignment system in place in San Francisco Unified School District has not supported the district's goals and values. We identify four values that the system was intended to support: (1) transparency, predictability and simplicity; (2) equity and diversity; (3) quality schools; and (4) community and continuity. We identify how the algorithm's theoretical promises to uphold these values depend on assumptions about how stakeholders behave and interact with the system, and explore the ways in which these assumptions clash with the properties and constraints of the real world. We discuss the implications of this work for algorithm design that accounts for complex and possibly conflicting values and needs.

\begin{acks}
We thank our study participants for sharing their experiences and insights. We also thank members of the U.C. Berkeley Algorithmic Fairness and Opacity Working Group (AFOG) and participants at the 4th Workshop on Mechanism Design for Social Good (MD4SG) for helpful feedback on an earlier version of this work. Finally, we thank the anonymous reviewers for their feedback and suggestions.
\end{acks}

\bibliographystyle{ACM-Reference-Format}
\bibliography{refs}


\begin{thebibliography}{82}


\ifx \showCODEN    \undefined \def \showCODEN     #1{\unskip}     \fi
\ifx \showDOI      \undefined \def \showDOI       #1{#1}\fi
\ifx \showISBNx    \undefined \def \showISBNx     #1{\unskip}     \fi
\ifx \showISBNxiii \undefined \def \showISBNxiii  #1{\unskip}     \fi
\ifx \showISSN     \undefined \def \showISSN      #1{\unskip}     \fi
\ifx \showLCCN     \undefined \def \showLCCN      #1{\unskip}     \fi
\ifx \shownote     \undefined \def \shownote      #1{#1}          \fi
\ifx \showarticletitle \undefined \def \showarticletitle #1{#1}   \fi
\ifx \showURL      \undefined \def \showURL       {\relax}        \fi
\providecommand\bibfield[2]{#2}
\providecommand\bibinfo[2]{#2}
\providecommand\natexlab[1]{#1}
\providecommand\showeprint[2][]{arXiv:#2}

\bibitem[\protect\citeauthoryear{Abdulkadiroglu, Pathak, Schellenberg, and
  Walters}{Abdulkadiroglu et~al\mbox{.}}{2017}]%
        {Abdulkadiroglu2019}
\bibfield{author}{\bibinfo{person}{Atila Abdulkadiroglu},
  \bibinfo{person}{Parag~A Pathak}, \bibinfo{person}{Jonathan Schellenberg},
  {and} \bibinfo{person}{Christopher~R Walters}.}
  \bibinfo{year}{2017}\natexlab{}.
\newblock \bibinfo{booktitle}{\emph{Do Parents Value School Effectiveness?}}
\newblock \bibinfo{type}{Working Paper} 23912. \bibinfo{institution}{National
  Bureau of Economic Research}.
\newblock
\urldef\tempurl%
\url{https://doi.org/10.3386/w23912}
\showDOI{\tempurl}


\bibitem[\protect\citeauthoryear{Abdulkadiroğlu and Sönmez}{Abdulkadiroğlu
  and Sönmez}{2003}]%
        {Abdulkadiroglu2003}
\bibfield{author}{\bibinfo{person}{Atila Abdulkadiroğlu} {and}
  \bibinfo{person}{Tayfun Sönmez}.} \bibinfo{year}{2003}\natexlab{}.
\newblock \showarticletitle{School Choice: A Mechanism Design Approach}.
\newblock \bibinfo{journal}{\emph{American Economic Review}}
  \bibinfo{volume}{93}, \bibinfo{number}{3} (\bibinfo{date}{June}
  \bibinfo{year}{2003}), \bibinfo{pages}{729--747}.
\newblock
\urldef\tempurl%
\url{https://doi.org/10.1257/000282803322157061}
\showDOI{\tempurl}


\bibitem[\protect\citeauthoryear{{African American Parent Advisory
  Council}}{{African American Parent Advisory Council}}{2017}]%
        {aapac}
\bibfield{author}{\bibinfo{person}{{African American Parent Advisory
  Council}}.} \bibinfo{year}{2017}\natexlab{}.
\newblock \bibinfo{booktitle}{\emph{AAPAC Reflections on SFUSD's Student
  Assignment Policy}}.
\newblock San Francisco Unified School District.
\newblock
\urldef\tempurl%
\url{https://archive.sfusd.edu/en/assets/sfusd-staff/enroll/files/AAPAC_Student_Assignment_Presentation_3.8.17.pdf?_ga=2.81786516.225077105.1586807928-1027271765.1579115407}
\showURL{%
Retrieved April 17, 2020 from \tempurl}


\bibitem[\protect\citeauthoryear{Ali, Sapiezynski, Bogen, Korolova, Mislove,
  and Rieke}{Ali et~al\mbox{.}}{2019}]%
        {Ali2019}
\bibfield{author}{\bibinfo{person}{Muhammad Ali}, \bibinfo{person}{Piotr
  Sapiezynski}, \bibinfo{person}{Miranda Bogen}, \bibinfo{person}{Aleksandra
  Korolova}, \bibinfo{person}{Alan Mislove}, {and} \bibinfo{person}{Aaron
  Rieke}.} \bibinfo{year}{2019}\natexlab{}.
\newblock \showarticletitle{Discrimination through Optimization: How Facebook's
  Ad Delivery Can Lead to Biased Outcomes}.
\newblock \bibinfo{journal}{\emph{Proc. ACM Hum.-Comput. Interact.}}
  \bibinfo{volume}{3}, \bibinfo{number}{CSCW}, Article \bibinfo{articleno}{199}
  (\bibinfo{date}{Nov.} \bibinfo{year}{2019}), \bibinfo{numpages}{30}~pages.
\newblock
\urldef\tempurl%
\url{https://doi.org/10.1145/3359301}
\showDOI{\tempurl}


\bibitem[\protect\citeauthoryear{Ames}{Ames}{2016}]%
        {Ames2016}
\bibfield{author}{\bibinfo{person}{Morgan~G. Ames}.}
  \bibinfo{year}{2016}\natexlab{}.
\newblock \showarticletitle{Learning consumption: Media, literacy, and the
  legacy of One Laptop per Child}.
\newblock \bibinfo{journal}{\emph{The Information Society}}
  \bibinfo{volume}{32}, \bibinfo{number}{2} (\bibinfo{year}{2016}),
  \bibinfo{pages}{85--97}.
\newblock
\urldef\tempurl%
\url{https://doi.org/10.1080/01972243.2016.1130497}
\showDOI{\tempurl}


\bibitem[\protect\citeauthoryear{Barnum and LeMee}{Barnum and LeMee}{2019}]%
        {barnum2019}
\bibfield{author}{\bibinfo{person}{Matt Barnum} {and}
  \bibinfo{person}{Gabrielle~LaMarr LeMee}.} \bibinfo{year}{2019}\natexlab{}.
\newblock \bibinfo{booktitle}{\emph{Looking for a home? You’ve seen
  GreatSchools ratings. Here’s how they nudge families toward schools with
  fewer black and Hispanic students.}}
\newblock Chalkbeat.
\newblock
\urldef\tempurl%
\url{https://www.chalkbeat.org/2019/12/5/21121858/looking-for-a-home-you-ve-seen-greatschools-ratings-here-s-how-they-nudge-families-toward-schools-wi}
\showURL{%
\tempurl}


\bibitem[\protect\citeauthoryear{Barry-Jester, Casselman, and
  Goldstein}{Barry-Jester et~al\mbox{.}}{2015}]%
        {BarryJester2015}
\bibfield{author}{\bibinfo{person}{Anna~Maria Barry-Jester},
  \bibinfo{person}{Ben Casselman}, {and} \bibinfo{person}{Dana Goldstein}.}
  \bibinfo{year}{2015}\natexlab{}.
\newblock \bibinfo{booktitle}{\emph{The New Science of Sentencing}}.
\newblock The Marshall Project.
\newblock
\urldef\tempurl%
\url{https://www.themarshallproject.org/2015/08/04/the-new-science-of-sentencing}
\showURL{%
\tempurl}


\bibitem[\protect\citeauthoryear{Baumer}{Baumer}{2017}]%
        {Baumer2017}
\bibfield{author}{\bibinfo{person}{Eric~PS Baumer}.}
  \bibinfo{year}{2017}\natexlab{}.
\newblock \showarticletitle{Toward human-centered algorithm design}.
\newblock \bibinfo{journal}{\emph{Big Data \& Society}} \bibinfo{volume}{4},
  \bibinfo{number}{2} (\bibinfo{year}{2017}), \bibinfo{pages}{1--12}.
\newblock
\urldef\tempurl%
\url{https://doi.org/10.1177/2053951717718854}
\showDOI{\tempurl}


\bibitem[\protect\citeauthoryear{Bolukbasi, Chang, Zou, Saligrama, and
  Kalai}{Bolukbasi et~al\mbox{.}}{2016}]%
        {Bolukbasi2016}
\bibfield{author}{\bibinfo{person}{Tolga Bolukbasi}, \bibinfo{person}{Kai-Wei
  Chang}, \bibinfo{person}{James Zou}, \bibinfo{person}{Venkatesh Saligrama},
  {and} \bibinfo{person}{Adam Kalai}.} \bibinfo{year}{2016}\natexlab{}.
\newblock \showarticletitle{Man is to Computer Programmer as Woman is to
  Homemaker? Debiasing Word Embeddings}. In
  \bibinfo{booktitle}{\emph{Proceedings of the 30th International Conference on
  Neural Information Processing Systems}} (Barcelona, Spain)
  \emph{(\bibinfo{series}{NIPS'16})}. \bibinfo{publisher}{Curran Associates
  Inc.}, \bibinfo{address}{Red Hook, NY, USA}, \bibinfo{pages}{4356–4364}.
\newblock
\showISBNx{9781510838819}


\bibitem[\protect\citeauthoryear{Borning and Muller}{Borning and
  Muller}{2012}]%
        {Borning2012}
\bibfield{author}{\bibinfo{person}{Alan Borning} {and} \bibinfo{person}{Michael
  Muller}.} \bibinfo{year}{2012}\natexlab{}.
\newblock \showarticletitle{Next Steps for Value Sensitive Design}. In
  \bibinfo{booktitle}{\emph{Proceedings of the SIGCHI Conference on Human
  Factors in Computing Systems}} (Austin, Texas, USA)
  \emph{(\bibinfo{series}{CHI ’12})}. \bibinfo{publisher}{Association for
  Computing Machinery}, \bibinfo{address}{New York, NY, USA},
  \bibinfo{pages}{1125–1134}.
\newblock
\showISBNx{9781450310154}
\urldef\tempurl%
\url{https://doi.org/10.1145/2207676.2208560}
\showDOI{\tempurl}


\bibitem[\protect\citeauthoryear{Box}{Box}{1979}]%
        {Box1979}
\bibfield{author}{\bibinfo{person}{G.~E.~P. Box}.}
  \bibinfo{year}{1979}\natexlab{}.
\newblock \showarticletitle{Robustness in the {Strategy} of {Scientific}
  {Model} {Building}}.
\newblock In \bibinfo{booktitle}{\emph{Robustness in {Statistics}}},
  \bibfield{editor}{\bibinfo{person}{Robert~L. Launder} {and}
  \bibinfo{person}{Graham~N. Wilkinson}} (Eds.). \bibinfo{publisher}{Academic
  Press}, \bibinfo{address}{Cambridge, MA, USA}, \bibinfo{pages}{201 -- 236}.
\newblock
\showISBNx{978-0-12-438150-6}
\urldef\tempurl%
\url{https://doi.org/10.1016/B978-0-12-438150-6.50018-2}
\showDOI{\tempurl}


\bibitem[\protect\citeauthoryear{Brown, Chouldechova, Putnam-Hornstein, Tobin,
  and Vaithianathan}{Brown et~al\mbox{.}}{2019}]%
        {Brown2019}
\bibfield{author}{\bibinfo{person}{Anna Brown}, \bibinfo{person}{Alexandra
  Chouldechova}, \bibinfo{person}{Emily Putnam-Hornstein},
  \bibinfo{person}{Andrew Tobin}, {and} \bibinfo{person}{Rhema Vaithianathan}.}
  \bibinfo{year}{2019}\natexlab{}.
\newblock \showarticletitle{Toward Algorithmic Accountability in Public
  Services: A Qualitative Study of Affected Community Perspectives on
  Algorithmic Decision-Making in Child Welfare Services}. In
  \bibinfo{booktitle}{\emph{Proceedings of the 2019 CHI Conference on Human
  Factors in Computing Systems}} (Glasgow, Scotland Uk)
  \emph{(\bibinfo{series}{CHI ’19})}. \bibinfo{publisher}{Association for
  Computing Machinery}, \bibinfo{address}{New York, NY, USA}, Article
  \bibinfo{articleno}{41}, \bibinfo{numpages}{12}~pages.
\newblock
\showISBNx{9781450359702}
\urldef\tempurl%
\url{https://doi.org/10.1145/3290605.3300271}
\showDOI{\tempurl}


\bibitem[\protect\citeauthoryear{Buolamwini and Gebru}{Buolamwini and
  Gebru}{2018}]%
        {Buolamwini2018}
\bibfield{author}{\bibinfo{person}{Joy Buolamwini} {and}
  \bibinfo{person}{Timnit Gebru}.} \bibinfo{year}{2018}\natexlab{}.
\newblock \showarticletitle{Gender Shades: Intersectional Accuracy Disparities
  in Commercial Gender Classification}. In
  \bibinfo{booktitle}{\emph{Proceedings of the 1st Conference on Fairness,
  Accountability and Transparency}} \emph{(\bibinfo{series}{Proceedings of
  Machine Learning Research})}, \bibfield{editor}{\bibinfo{person}{Sorelle~A.
  Friedler} {and} \bibinfo{person}{Christo Wilson}} (Eds.),
  Vol.~\bibinfo{volume}{81}. \bibinfo{publisher}{PMLR}, \bibinfo{address}{New
  York, NY, USA}, \bibinfo{pages}{77--91}.
\newblock
\urldef\tempurl%
\url{http://proceedings.mlr.press/v81/buolamwini18a.html}
\showURL{%
\tempurl}


\bibitem[\protect\citeauthoryear{Burgess, Greaves, Vignoles, and
  Wilson}{Burgess et~al\mbox{.}}{2015}]%
        {Burgess2015}
\bibfield{author}{\bibinfo{person}{Simon Burgess}, \bibinfo{person}{Ellen
  Greaves}, \bibinfo{person}{Anna Vignoles}, {and} \bibinfo{person}{Deborah
  Wilson}.} \bibinfo{year}{2015}\natexlab{}.
\newblock \showarticletitle{What Parents Want: School Preferences and School
  Choice}.
\newblock \bibinfo{journal}{\emph{The Economic Journal}} \bibinfo{volume}{125},
  \bibinfo{number}{587} (\bibinfo{year}{2015}), \bibinfo{pages}{1262--1289}.
\newblock
\urldef\tempurl%
\url{https://doi.org/10.1111/ecoj.12153}
\showDOI{\tempurl}


\bibitem[\protect\citeauthoryear{Chancellor, Baumer, and
  De~Choudhury}{Chancellor et~al\mbox{.}}{2019}]%
        {Chancellor2019}
\bibfield{author}{\bibinfo{person}{Stevie Chancellor}, \bibinfo{person}{Eric
  P.~S. Baumer}, {and} \bibinfo{person}{Munmun De~Choudhury}.}
  \bibinfo{year}{2019}\natexlab{}.
\newblock \showarticletitle{Who is the "Human" in Human-Centered Machine
  Learning: The Case of Predicting Mental Health from Social Media}.
\newblock \bibinfo{journal}{\emph{Proc. ACM Hum.-Comput. Interact.}}
  \bibinfo{volume}{3}, \bibinfo{number}{CSCW}, Article \bibinfo{articleno}{147}
  (\bibinfo{date}{Nov.} \bibinfo{year}{2019}), \bibinfo{numpages}{32}~pages.
\newblock
\urldef\tempurl%
\url{https://doi.org/10.1145/3359249}
\showDOI{\tempurl}


\bibitem[\protect\citeauthoryear{Charmaz}{Charmaz}{2014}]%
        {groundedtheory}
\bibfield{author}{\bibinfo{person}{Kathy Charmaz}.}
  \bibinfo{year}{2014}\natexlab{}.
\newblock \bibinfo{booktitle}{\emph{Constructing grounded theory: A practical
  guide through qualitative research} (\bibinfo{edition}{2} ed.)}.
\newblock \bibinfo{publisher}{SAGE Publications Ltd}, \bibinfo{address}{London,
  United Kingdom}.
\newblock


\bibitem[\protect\citeauthoryear{Chen and He}{Chen and He}{2020}]%
        {Chen2017}
\bibfield{author}{\bibinfo{person}{Yan Chen} {and} \bibinfo{person}{Yinghua
  He}.} \bibinfo{year}{2020}\natexlab{}.
\newblock \bibinfo{booktitle}{\emph{Information Acquisition and Provision in
  School Choice : An Experimental Study}}.
\newblock Working Paper.
\newblock
\urldef\tempurl%
\url{http://yanchen.people.si.umich.edu/papers/Chen_He_2020_09_Distribute.pdf}
\showURL{%
\tempurl}


\bibitem[\protect\citeauthoryear{Cheng, Wang, Zhang, O'Connell, Gray, Harper,
  and Zhu}{Cheng et~al\mbox{.}}{2019}]%
        {Cheng2019}
\bibfield{author}{\bibinfo{person}{Hao-Fei Cheng}, \bibinfo{person}{Ruotong
  Wang}, \bibinfo{person}{Zheng Zhang}, \bibinfo{person}{Fiona O'Connell},
  \bibinfo{person}{Terrance Gray}, \bibinfo{person}{F.~Maxwell Harper}, {and}
  \bibinfo{person}{Haiyi Zhu}.} \bibinfo{year}{2019}\natexlab{}.
\newblock \showarticletitle{Explaining Decision-Making Algorithms through UI:
  Strategies to Help Non-Expert Stakeholders}. In
  \bibinfo{booktitle}{\emph{Proceedings of the 2019 CHI Conference on Human
  Factors in Computing Systems}} (Glasgow, Scotland Uk)
  \emph{(\bibinfo{series}{CHI '19})}. \bibinfo{publisher}{Association for
  Computing Machinery}, \bibinfo{address}{New York, NY, USA},
  \bibinfo{pages}{1–12}.
\newblock
\showISBNx{9781450359702}
\urldef\tempurl%
\url{https://doi.org/10.1145/3290605.3300789}
\showDOI{\tempurl}


\bibitem[\protect\citeauthoryear{Chouldechova}{Chouldechova}{2017}]%
        {Chouldechova2017}
\bibfield{author}{\bibinfo{person}{Alexandra Chouldechova}.}
  \bibinfo{year}{2017}\natexlab{}.
\newblock \showarticletitle{Fair Prediction with Disparate Impact: A Study of
  Bias in Recidivism Prediction Instruments}.
\newblock \bibinfo{journal}{\emph{Big Data}} \bibinfo{volume}{5},
  \bibinfo{number}{2} (\bibinfo{year}{2017}), \bibinfo{pages}{153--163}.
\newblock
\urldef\tempurl%
\url{https://doi.org/10.1089/big.2016.0047}
\showDOI{\tempurl}
\newblock
\shownote{PMID: 28632438.}


\bibitem[\protect\citeauthoryear{Cifor, Garcia, Cowan, Rault, Sutherland, Chan,
  Rode, Hoffmann, Salehi, and Nakamura}{Cifor et~al\mbox{.}}{2019}]%
        {cifor2019feminist}
\bibfield{author}{\bibinfo{person}{Marika Cifor}, \bibinfo{person}{Patricia
  Garcia}, \bibinfo{person}{TL Cowan}, \bibinfo{person}{Jasmine Rault},
  \bibinfo{person}{Tonia Sutherland}, \bibinfo{person}{Anita~Say Chan},
  \bibinfo{person}{Jennifer Rode}, \bibinfo{person}{Anna~Lauren Hoffmann},
  \bibinfo{person}{Niloufar Salehi}, {and} \bibinfo{person}{Lisa Nakamura}.}
  \bibinfo{year}{2019}\natexlab{}.
\newblock \bibinfo{title}{Feminist data manifest-no}.
\newblock
\newblock
\urldef\tempurl%
\url{https://www.manifestno.com/}
\showURL{%
\tempurl}


\bibitem[\protect\citeauthoryear{Costanza-Chock}{Costanza-Chock}{2018}]%
        {costanza2018design}
\bibfield{author}{\bibinfo{person}{Sasha Costanza-Chock}.}
  \bibinfo{year}{2018}\natexlab{}.
\newblock \showarticletitle{Design Justice: Towards an Intersectional Feminist
  Framework for Design Theory and Practice}. In
  \bibinfo{booktitle}{\emph{Proceedings of the Design Research Society}}.
  \bibinfo{publisher}{Design Research Society}, \bibinfo{address}{London,
  United Kingdom}, 14.
\newblock
\urldef\tempurl%
\url{https://ssrn.com/abstract=3189696}
\showURL{%
\tempurl}


\bibitem[\protect\citeauthoryear{Crawford}{Crawford}{2016}]%
        {Crawford2016}
\bibfield{author}{\bibinfo{person}{Kate Crawford}.}
  \bibinfo{year}{2016}\natexlab{}.
\newblock \showarticletitle{Can an Algorithm be Agonistic? Ten Scenes from Life
  in Calculated Publics}.
\newblock \bibinfo{journal}{\emph{Science, Technology, \& Human Values}}
  \bibinfo{volume}{41}, \bibinfo{number}{1} (\bibinfo{year}{2016}),
  \bibinfo{pages}{77--92}.
\newblock
\urldef\tempurl%
\url{https://doi.org/10.1177/0162243915589635}
\showDOI{\tempurl}


\bibitem[\protect\citeauthoryear{Davis and Nathan}{Davis and Nathan}{2015}]%
        {Davis2015}
\bibfield{author}{\bibinfo{person}{Janet Davis} {and} \bibinfo{person}{Lisa~P.
  Nathan}.} \bibinfo{year}{2015}\natexlab{}.
\newblock \showarticletitle{Value Sensitive Design: Applications, Adaptations,
  and Critiques}.
\newblock In \bibinfo{booktitle}{\emph{Handbook of Ethics, Values, and
  Technological Design}}, \bibfield{editor}{\bibinfo{person}{Jeroen van~den
  Hoven}, \bibinfo{person}{Pieter~E. Vermaas}, {and} \bibinfo{person}{Ibo
  van~de Poel}} (Eds.). \bibinfo{publisher}{Springer},
  \bibinfo{address}{Dordrecht, Netherlands}, \bibinfo{pages}{11--40}.
\newblock


\bibitem[\protect\citeauthoryear{Dietz, Ostrom, and Stern}{Dietz
  et~al\mbox{.}}{2003}]%
        {Dietz2003}
\bibfield{author}{\bibinfo{person}{Thomas Dietz}, \bibinfo{person}{Elinor
  Ostrom}, {and} \bibinfo{person}{Paul~C. Stern}.}
  \bibinfo{year}{2003}\natexlab{}.
\newblock \showarticletitle{The Struggle to Govern the Commons}.
\newblock \bibinfo{journal}{\emph{Science}} \bibinfo{volume}{302},
  \bibinfo{number}{5652} (\bibinfo{year}{2003}), \bibinfo{pages}{1907--1912}.
\newblock
\showISSN{0036-8075}
\urldef\tempurl%
\url{https://doi.org/10.1126/science.1091015}
\showDOI{\tempurl}


\bibitem[\protect\citeauthoryear{Ding and Schotter}{Ding and Schotter}{2017}]%
        {Ding2017}
\bibfield{author}{\bibinfo{person}{Tingting Ding} {and} \bibinfo{person}{Andrew
  Schotter}.} \bibinfo{year}{2017}\natexlab{}.
\newblock \showarticletitle{Matching and chatting: An experimental study of the
  impact of network communication on school-matching mechanisms}.
\newblock \bibinfo{journal}{\emph{Games and Economic Behavior}}
  \bibinfo{volume}{103} (\bibinfo{year}{2017}), \bibinfo{pages}{94 -- 115}.
\newblock
\showISSN{0899-8256}
\urldef\tempurl%
\url{https://doi.org/10.1016/j.geb.2016.02.004}
\showDOI{\tempurl}
\newblock
\shownote{John Nash Memorial.}


\bibitem[\protect\citeauthoryear{Ding and Schotter}{Ding and Schotter}{2019}]%
        {Ding2019}
\bibfield{author}{\bibinfo{person}{Tingting Ding} {and} \bibinfo{person}{Andrew
  Schotter}.} \bibinfo{year}{2019}\natexlab{}.
\newblock \showarticletitle{{Learning and Mechanism Design: An Experimental
  Test of School Matching Mechanisms with Intergenerational Advice}}.
\newblock \bibinfo{journal}{\emph{The Economic Journal}} \bibinfo{volume}{129},
  \bibinfo{number}{623} (\bibinfo{date}{05} \bibinfo{year}{2019}),
  \bibinfo{pages}{2779--2804}.
\newblock


\bibitem[\protect\citeauthoryear{Efron and Tibshirani}{Efron and
  Tibshirani}{1993}]%
        {bootstrap}
\bibfield{author}{\bibinfo{person}{Bradley Efron} {and}
  \bibinfo{person}{Robert~J. Tibshirani}.} \bibinfo{year}{1993}\natexlab{}.
\newblock \bibinfo{booktitle}{\emph{An Introduction to the Bootstrap}}.
\newblock \bibinfo{publisher}{Chapman \& Hall/CRC}, \bibinfo{address}{Boca
  Raton, FL, USA}.
\newblock


\bibitem[\protect\citeauthoryear{Ewing}{Ewing}{2018}]%
        {ewing2018}
\bibfield{author}{\bibinfo{person}{Eve~L. Ewing}.}
  \bibinfo{year}{2018}\natexlab{}.
\newblock \bibinfo{booktitle}{\emph{Ghosts in the Schoolyard: Racism and School
  Closings on Chicago's South Side}}.
\newblock \bibinfo{publisher}{University of Chicago Press},
  \bibinfo{address}{Chicago, IL, USA}.
\newblock


\bibitem[\protect\citeauthoryear{Flanagan, Howe, and Nissenbaum}{Flanagan
  et~al\mbox{.}}{2005}]%
        {Flanagan2005}
\bibfield{author}{\bibinfo{person}{Mary Flanagan}, \bibinfo{person}{Daniel~C.
  Howe}, {and} \bibinfo{person}{Helen Nissenbaum}.}
  \bibinfo{year}{2005}\natexlab{}.
\newblock \showarticletitle{Values at Play: Design Tradeoffs in
  Socially-Oriented Game Design}. In \bibinfo{booktitle}{\emph{Proceedings of
  the SIGCHI Conference on Human Factors in Computing Systems}} (Portland,
  Oregon, USA) \emph{(\bibinfo{series}{CHI '05})}.
  \bibinfo{publisher}{Association for Computing Machinery},
  \bibinfo{address}{New York, NY, USA}, \bibinfo{pages}{751–760}.
\newblock
\showISBNx{1581139985}
\urldef\tempurl%
\url{https://doi.org/10.1145/1054972.1055076}
\showDOI{\tempurl}


\bibitem[\protect\citeauthoryear{Friedman, Hendry, and Borning}{Friedman
  et~al\mbox{.}}{2017}]%
        {Friedman2017}
\bibfield{author}{\bibinfo{person}{Batya Friedman}, \bibinfo{person}{David~G.
  Hendry}, {and} \bibinfo{person}{Alan Borning}.}
  \bibinfo{year}{2017}\natexlab{}.
\newblock \showarticletitle{A Survey of Value Sensitive Design Methods}.
\newblock \bibinfo{journal}{\emph{Foundations and Trends in Human–Computer
  Interaction}} \bibinfo{volume}{11}, \bibinfo{number}{2}
  (\bibinfo{year}{2017}), \bibinfo{pages}{63--125}.
\newblock
\showISSN{1551-3955}
\urldef\tempurl%
\url{https://doi.org/10.1561/1100000015}
\showDOI{\tempurl}


\bibitem[\protect\citeauthoryear{Friedman and Jr.}{Friedman and Jr.}{2003}]%
        {Friedman2003}
\bibfield{author}{\bibinfo{person}{Batya Friedman} {and} \bibinfo{person}{Peter
  H.~Kahn Jr.}} \bibinfo{year}{2003}\natexlab{}.
\newblock \showarticletitle{Human Values, Ethics, and Design}.
\newblock In \bibinfo{booktitle}{\emph{The Human-Computer Interaction Handbook:
  Fundamentals, Evolving Technologies, and Emerging Applications}},
  \bibfield{editor}{\bibinfo{person}{Jacko~JA Sears~A}} (Ed.).
  \bibinfo{publisher}{L. Erlbaum Associates Inc.}, \bibinfo{address}{USA},
  \bibinfo{pages}{1177--1201}.
\newblock


\bibitem[\protect\citeauthoryear{Friedman, Kahn, and Borning}{Friedman
  et~al\mbox{.}}{2006}]%
        {Friedman2006}
\bibfield{author}{\bibinfo{person}{B Friedman}, \bibinfo{person}{PH~Jr Kahn},
  {and} \bibinfo{person}{A Borning}.} \bibinfo{year}{2006}\natexlab{}.
\newblock \showarticletitle{Value Sensitive Design and Information Systems}.
\newblock In \bibinfo{booktitle}{\emph{Human-Computer Interaction in Management
  Information Systems: Foundations}}, \bibfield{editor}{\bibinfo{person}{Ben
  Shneiderman}, \bibinfo{person}{Ping Zhang}, {and} \bibinfo{person}{Dennis
  Galletta}} (Eds.). \bibinfo{publisher}{M. E. Sharpe, Inc.},
  \bibinfo{address}{Armonk, NY, USA}, \bibinfo{pages}{348--372}.
\newblock


\bibitem[\protect\citeauthoryear{Gale and Shapley}{Gale and Shapley}{1962}]%
        {GaleShapley1962}
\bibfield{author}{\bibinfo{person}{D. Gale} {and} \bibinfo{person}{L.~S.
  Shapley}.} \bibinfo{year}{1962}\natexlab{}.
\newblock \showarticletitle{College Admissions and the Stability of Marriage}.
\newblock \bibinfo{journal}{\emph{The American Mathematical Monthly}}
  \bibinfo{volume}{69}, \bibinfo{number}{2} (\bibinfo{date}{1}
  \bibinfo{year}{1962}), \bibinfo{pages}{9--15}.
\newblock


\bibitem[\protect\citeauthoryear{Ghassemi, Naumann, Schulam, Beam, Chen, and
  Ranganath}{Ghassemi et~al\mbox{.}}{2020}]%
        {Ghassemi2019}
\bibfield{author}{\bibinfo{person}{Marzyeh Ghassemi}, \bibinfo{person}{Tristan
  Naumann}, \bibinfo{person}{Peter Schulam}, \bibinfo{person}{Andrew~L. Beam},
  \bibinfo{person}{Irene Ya-Ping Chen}, {and} \bibinfo{person}{Rajesh
  Ranganath}.} \bibinfo{year}{2020}\natexlab{}.
\newblock \showarticletitle{A Review of Challenges and Opportunities in Machine
  Learning for Health.}. In \bibinfo{booktitle}{\emph{AMIA Joint Summits on
  Translational Science}}. \bibinfo{publisher}{American Medical Informatics
  Association}, \bibinfo{address}{USA}, \bibinfo{pages}{191--200}.
\newblock


\bibitem[\protect\citeauthoryear{Glazerman and Dotter}{Glazerman and
  Dotter}{2017}]%
        {Glazerman2017}
\bibfield{author}{\bibinfo{person}{Steven Glazerman} {and}
  \bibinfo{person}{Dallas Dotter}.} \bibinfo{year}{2017}\natexlab{}.
\newblock \showarticletitle{Market Signals: Evidence on the Determinants and
  Consequences of School Choice From a Citywide Lottery}.
\newblock \bibinfo{journal}{\emph{Educational Evaluation and Policy Analysis}}
  \bibinfo{volume}{39}, \bibinfo{number}{4} (\bibinfo{year}{2017}),
  \bibinfo{pages}{593--619}.
\newblock
\urldef\tempurl%
\url{https://doi.org/10.3102/0162373717702964}
\showDOI{\tempurl}


\bibitem[\protect\citeauthoryear{Gonczarowski, Nisan, Kovalio, and
  Romm}{Gonczarowski et~al\mbox{.}}{2019}]%
        {Gonczarowski2019}
\bibfield{author}{\bibinfo{person}{Yannai~A. Gonczarowski},
  \bibinfo{person}{Noam Nisan}, \bibinfo{person}{Lior Kovalio}, {and}
  \bibinfo{person}{Assaf Romm}.} \bibinfo{year}{2019}\natexlab{}.
\newblock \showarticletitle{Matching for the Israeli "Mechinot" Gap-Year
  Programs: Handling Rich Diversity Requirements}. In
  \bibinfo{booktitle}{\emph{Proceedings of the 2019 ACM Conference on Economics
  and Computation}} (Phoenix, AZ, USA) \emph{(\bibinfo{series}{EC '19})}.
  \bibinfo{publisher}{Association for Computing Machinery},
  \bibinfo{address}{New York, NY, USA}, \bibinfo{pages}{321}.
\newblock
\showISBNx{9781450367929}
\urldef\tempurl%
\url{https://doi.org/10.1145/3328526.3329620}
\showDOI{\tempurl}


\bibitem[\protect\citeauthoryear{Griffith and Freedman}{Griffith and
  Freedman}{2019a}]%
        {schoolcolors}
\bibfield{author}{\bibinfo{person}{Mark~Winston Griffith} {and}
  \bibinfo{person}{Max Freedman}.} \bibinfo{year}{2019}\natexlab{a}.
\newblock \bibinfo{title}{School Colors (Episode 5: The Disappearing
  District)}.
\newblock
\newblock
\urldef\tempurl%
\url{https://www.schoolcolorspodcast.com/episodes/episode-5-the-disappearing-district}
\showURL{%
\tempurl}


\bibitem[\protect\citeauthoryear{Griffith and Freedman}{Griffith and
  Freedman}{2019b}]%
        {schoolcolors2}
\bibfield{author}{\bibinfo{person}{Mark~Winston Griffith} {and}
  \bibinfo{person}{Max Freedman}.} \bibinfo{year}{2019}\natexlab{b}.
\newblock \bibinfo{title}{School Colors (Episode 7: New Kids on the Block)}.
\newblock
\newblock
\urldef\tempurl%
\url{https://www.schoolcolorspodcast.com/episodes/episode-7-new-kids-on-the-block}
\showURL{%
\tempurl}


\bibitem[\protect\citeauthoryear{Guillen and Hakimov}{Guillen and
  Hakimov}{2017}]%
        {Guillen2017}
\bibfield{author}{\bibinfo{person}{Pablo Guillen} {and}
  \bibinfo{person}{Rustamdjan Hakimov}.} \bibinfo{year}{2017}\natexlab{}.
\newblock \showarticletitle{Not quite the best response: truth-telling,
  strategy-proof matching, and the manipulation of others}.
\newblock \bibinfo{journal}{\emph{Experimental Economics}}
  \bibinfo{volume}{20}, \bibinfo{number}{3} (\bibinfo{year}{2017}),
  \bibinfo{pages}{670--686}.
\newblock


\bibitem[\protect\citeauthoryear{Guillen and Hakimov}{Guillen and
  Hakimov}{2018}]%
        {Guillen2018}
\bibfield{author}{\bibinfo{person}{Pablo Guillen} {and}
  \bibinfo{person}{Rustamdjan Hakimov}.} \bibinfo{year}{2018}\natexlab{}.
\newblock \showarticletitle{The effectiveness of top-down advice in
  strategy-proof mechanisms: A field experiment}.
\newblock \bibinfo{journal}{\emph{European Economic Review}}
  \bibinfo{volume}{101}, \bibinfo{number}{C} (\bibinfo{year}{2018}),
  \bibinfo{pages}{505--511}.
\newblock


\bibitem[\protect\citeauthoryear{Guillen and Hing}{Guillen and Hing}{2014}]%
        {Guillen2014}
\bibfield{author}{\bibinfo{person}{Pablo Guillen} {and}
  \bibinfo{person}{Alexander Hing}.} \bibinfo{year}{2014}\natexlab{}.
\newblock \showarticletitle{Lying through their teeth: Third party advice and
  truth telling in a strategy proof mechanism}.
\newblock \bibinfo{journal}{\emph{European Economic Review}}
  \bibinfo{volume}{70} (\bibinfo{year}{2014}), \bibinfo{pages}{178 -- 185}.
\newblock
\showISSN{0014-2921}
\urldef\tempurl%
\url{https://doi.org/10.1016/j.euroecorev.2014.05.002}
\showDOI{\tempurl}


\bibitem[\protect\citeauthoryear{Hafalir, Yenmez, and Yildirim}{Hafalir
  et~al\mbox{.}}{2013}]%
        {Hafalir2013}
\bibfield{author}{\bibinfo{person}{Isa~E. Hafalir}, \bibinfo{person}{M.~Bumin
  Yenmez}, {and} \bibinfo{person}{Muhammed~A. Yildirim}.}
  \bibinfo{year}{2013}\natexlab{}.
\newblock \showarticletitle{Effective affirmative action in school choice}.
\newblock \bibinfo{journal}{\emph{Theoretical Economics}} \bibinfo{volume}{8},
  \bibinfo{number}{2} (\bibinfo{year}{2013}), \bibinfo{pages}{325--363}.
\newblock
\urldef\tempurl%
\url{https://doi.org/10.3982/TE1135}
\showDOI{\tempurl}


\bibitem[\protect\citeauthoryear{Haney, Cook, and Norton}{Haney
  et~al\mbox{.}}{2018}]%
        {sfusd2018}
\bibfield{author}{\bibinfo{person}{Matt Haney}, \bibinfo{person}{Stevon Cook},
  {and} \bibinfo{person}{Rachel Norton}.} \bibinfo{year}{2018}\natexlab{}.
\newblock \bibinfo{booktitle}{\emph{Developing a Community Based Student
  Assignment System for SFUSD}}.
\newblock San Francisco Unified School District.
\newblock
\urldef\tempurl%
\url{https://archive.sfusd.edu/en/assets/sfusd-staff/enroll/files/2019-20/Student\%20Assignment\%20Proposal\%20189-25A1.pdf}
\showURL{%
Retrieved April 13, 2020 from \tempurl}


\bibitem[\protect\citeauthoryear{Harrington, Erete, and Piper}{Harrington
  et~al\mbox{.}}{2019}]%
        {Harrington2019}
\bibfield{author}{\bibinfo{person}{Christina Harrington},
  \bibinfo{person}{Sheena Erete}, {and} \bibinfo{person}{Anne~Marie Piper}.}
  \bibinfo{year}{2019}\natexlab{}.
\newblock \showarticletitle{Deconstructing Community-Based Collaborative
  Design: Towards More Equitable Participatory Design Engagements}.
\newblock \bibinfo{journal}{\emph{Proc. ACM Hum.-Comput. Interact.}}
  \bibinfo{volume}{3}, \bibinfo{number}{CSCW}, Article \bibinfo{articleno}{216}
  (\bibinfo{date}{Nov.} \bibinfo{year}{2019}), \bibinfo{numpages}{25}~pages.
\newblock
\urldef\tempurl%
\url{https://doi.org/10.1145/3359318}
\showDOI{\tempurl}


\bibitem[\protect\citeauthoryear{Hassidim, Romm, and Shorrer}{Hassidim
  et~al\mbox{.}}{2016}]%
        {Hassidim2016}
\bibfield{author}{\bibinfo{person}{Avinatan Hassidim}, \bibinfo{person}{Assaf
  Romm}, {and} \bibinfo{person}{Ran~I. Shorrer}.}
  \bibinfo{year}{2016}\natexlab{}.
\newblock \showarticletitle{“Strategic” Behavior in a Strategy-Proof
  Environment}. In \bibinfo{booktitle}{\emph{Proceedings of the 2016 ACM
  Conference on Economics and Computation}} (Maastricht, The Netherlands)
  \emph{(\bibinfo{series}{EC ’16})}. \bibinfo{publisher}{Association for
  Computing Machinery}, \bibinfo{address}{New York, NY, USA},
  \bibinfo{pages}{763–764}.
\newblock
\showISBNx{9781450339360}
\urldef\tempurl%
\url{https://doi.org/10.1145/2940716.2940751}
\showDOI{\tempurl}


\bibitem[\protect\citeauthoryear{Hastings, Kane, and Staiger}{Hastings
  et~al\mbox{.}}{2007}]%
        {Hastings2009}
\bibfield{author}{\bibinfo{person}{Justine~S. Hastings},
  \bibinfo{person}{Thomas~J. Kane}, {and} \bibinfo{person}{Douglas~O.
  Staiger}.} \bibinfo{year}{2007}\natexlab{}.
\newblock \bibinfo{booktitle}{\emph{Heterogeneous Preferences and the Efficacy
  of Public School Choice}}.
\newblock \bibinfo{type}{Working Paper} 12145. \bibinfo{institution}{National
  Bureau of Economic Research}.
\newblock
\urldef\tempurl%
\url{https://doi.org/10.3386/w12145}
\showDOI{\tempurl}


\bibitem[\protect\citeauthoryear{Hermstrüwer}{Hermstrüwer}{2019}]%
        {Hermstruwer2019}
\bibfield{author}{\bibinfo{person}{Yoan Hermstrüwer}.}
  \bibinfo{year}{2019}\natexlab{}.
\newblock \bibinfo{booktitle}{\emph{Transparency and Fairness in School Choice
  Mechanisms}}.
\newblock \bibinfo{type}{{T}echnical {R}eport}. \bibinfo{institution}{Max
  Planck Institute for Research on Collective Goods}.
\newblock


\bibitem[\protect\citeauthoryear{Hitzig}{Hitzig}{2020}]%
        {Hitzig2019}
\bibfield{author}{\bibinfo{person}{Zoë Hitzig}.}
  \bibinfo{year}{2020}\natexlab{}.
\newblock \showarticletitle{The normative gap: Mechanism design and ideal
  theories of justice}.
\newblock \bibinfo{journal}{\emph{Economics and Philosophy}}
  \bibinfo{volume}{36}, \bibinfo{number}{3} (\bibinfo{year}{2020}),
  \bibinfo{pages}{407--434}.
\newblock
\urldef\tempurl%
\url{https://doi.org/10.1017/S0266267119000270}
\showDOI{\tempurl}


\bibitem[\protect\citeauthoryear{Kapor, Neilson, and Zimmerman}{Kapor
  et~al\mbox{.}}{2020}]%
        {Kapor2020}
\bibfield{author}{\bibinfo{person}{Adam~J. Kapor},
  \bibinfo{person}{Christopher~A. Neilson}, {and} \bibinfo{person}{Seth~D.
  Zimmerman}.} \bibinfo{year}{2020}\natexlab{}.
\newblock \showarticletitle{Heterogeneous Beliefs and School Choice
  Mechanisms}.
\newblock \bibinfo{journal}{\emph{American Economic Review}}
  \bibinfo{volume}{110}, \bibinfo{number}{5} (\bibinfo{date}{May}
  \bibinfo{year}{2020}), \bibinfo{pages}{1274--1315}.
\newblock
\urldef\tempurl%
\url{https://doi.org/10.1257/aer.20170129}
\showDOI{\tempurl}


\bibitem[\protect\citeauthoryear{Kasman and Valant}{Kasman and Valant}{2019}]%
        {Kasman2019}
\bibfield{author}{\bibinfo{person}{Matt Kasman} {and} \bibinfo{person}{Jon
  Valant}.} \bibinfo{year}{2019}\natexlab{}.
\newblock \bibinfo{booktitle}{\emph{The opportunities and risks of K-12 student
  placement algorithms}}.
\newblock The Brookings Institute.
\newblock
\urldef\tempurl%
\url{https://www.brookings.edu/research/the-opportunities-and-risks-of-k-12-student-placement-algorithms/}
\showURL{%
\tempurl}


\bibitem[\protect\citeauthoryear{Koren}{Koren}{2015}]%
        {Koren2015}
\bibfield{author}{\bibinfo{person}{James~Rufus Koren}.}
  \bibinfo{year}{2015}\natexlab{}.
\newblock \bibinfo{booktitle}{\emph{Some lenders are judging you on much more
  than finances}}.
\newblock Los Angeles Times.
\newblock
\urldef\tempurl%
\url{https://www.latimes.com/business/la-fi-new-credit-score-20151220-story.html}
\showURL{%
\tempurl}


\bibitem[\protect\citeauthoryear{{Kulesza}, {Stumpf}, {Burnett}, {Yang},
  {Kwan}, and {Wong}}{{Kulesza} et~al\mbox{.}}{2013}]%
        {Kulesza2013}
\bibfield{author}{\bibinfo{person}{T. {Kulesza}}, \bibinfo{person}{S.
  {Stumpf}}, \bibinfo{person}{M. {Burnett}}, \bibinfo{person}{S. {Yang}},
  \bibinfo{person}{I. {Kwan}}, {and} \bibinfo{person}{W. {Wong}}.}
  \bibinfo{year}{2013}\natexlab{}.
\newblock \showarticletitle{Too much, too little, or just right? Ways
  explanations impact end users' mental models}. In
  \bibinfo{booktitle}{\emph{2013 IEEE Symposium on Visual Languages and Human
  Centric Computing}}. \bibinfo{publisher}{IEEE}, \bibinfo{address}{San Jose,
  CA, USA}, \bibinfo{pages}{3--10}.
\newblock
\urldef\tempurl%
\url{https://doi.org/10.1109/VLHCC.2013.6645235}
\showDOI{\tempurl}


\bibitem[\protect\citeauthoryear{Laverde}{Laverde}{2020}]%
        {Laverde2020}
\bibfield{author}{\bibinfo{person}{Mariana Laverde}.}
  \bibinfo{year}{2020}\natexlab{}.
\newblock \bibinfo{booktitle}{\emph{Unequal Assignments to Public Schools and
  the Limits of School Choice}}.
\newblock Working Paper.
\newblock
\urldef\tempurl%
\url{https://drive.google.com/file/d/19HJFGmaf2HA56k3rNBgB_at4aSkG3Bpz/view?usp=sharing}
\showURL{%
\tempurl}


\bibitem[\protect\citeauthoryear{Le~Dantec, Poole, and Wyche}{Le~Dantec
  et~al\mbox{.}}{2009}]%
        {LeDantec2009}
\bibfield{author}{\bibinfo{person}{Christopher~A. Le~Dantec},
  \bibinfo{person}{Erika~Shehan Poole}, {and} \bibinfo{person}{Susan~P.
  Wyche}.} \bibinfo{year}{2009}\natexlab{}.
\newblock \showarticletitle{Values as Lived Experience: Evolving Value
  Sensitive Design in Support of Value Discovery}. In
  \bibinfo{booktitle}{\emph{Proceedings of the SIGCHI Conference on Human
  Factors in Computing Systems}} (Boston, MA, USA) \emph{(\bibinfo{series}{CHI
  '09})}. \bibinfo{publisher}{Association for Computing Machinery},
  \bibinfo{address}{New York, NY, USA}, \bibinfo{pages}{1141–1150}.
\newblock
\showISBNx{9781605582467}
\urldef\tempurl%
\url{https://doi.org/10.1145/1518701.1518875}
\showURL{%
\tempurl}


\bibitem[\protect\citeauthoryear{Lee and Baykal}{Lee and Baykal}{2017}]%
        {Lee2017a}
\bibfield{author}{\bibinfo{person}{Min~Kyung Lee} {and} \bibinfo{person}{Su
  Baykal}.} \bibinfo{year}{2017}\natexlab{}.
\newblock \showarticletitle{Algorithmic Mediation in Group Decisions: Fairness
  Perceptions of Algorithmically Mediated vs. Discussion-Based Social
  Division}. In \bibinfo{booktitle}{\emph{Proceedings of the 2017 ACM
  Conference on Computer Supported Cooperative Work and Social Computing}}
  (Portland, Oregon, USA) \emph{(\bibinfo{series}{CSCW ’17})}.
  \bibinfo{publisher}{Association for Computing Machinery},
  \bibinfo{address}{New York, NY, USA}, \bibinfo{pages}{1035–1048}.
\newblock
\showISBNx{9781450343350}
\urldef\tempurl%
\url{https://doi.org/10.1145/2998181.2998230}
\showDOI{\tempurl}


\bibitem[\protect\citeauthoryear{Lee, Jain, Cha, Ojha, and Kusbit}{Lee
  et~al\mbox{.}}{2019}]%
        {Lee2019}
\bibfield{author}{\bibinfo{person}{Min~Kyung Lee}, \bibinfo{person}{Anuraag
  Jain}, \bibinfo{person}{Hea~Jin Cha}, \bibinfo{person}{Shashank Ojha}, {and}
  \bibinfo{person}{Daniel Kusbit}.} \bibinfo{year}{2019}\natexlab{}.
\newblock \showarticletitle{Procedural Justice in Algorithmic Fairness:
  Leveraging Transparency and Outcome Control for Fair Algorithmic Mediation}.
\newblock \bibinfo{journal}{\emph{Proc. ACM Hum.-Comput. Interact.}}
  \bibinfo{volume}{3}, \bibinfo{number}{CSCW}, Article \bibinfo{articleno}{182}
  (\bibinfo{date}{Nov.} \bibinfo{year}{2019}), \bibinfo{numpages}{26}~pages.
\newblock
\urldef\tempurl%
\url{https://doi.org/10.1145/3359284}
\showDOI{\tempurl}


\bibitem[\protect\citeauthoryear{Lee, Kim, and Lizarondo}{Lee
  et~al\mbox{.}}{2017}]%
        {Lee2017b}
\bibfield{author}{\bibinfo{person}{Min~Kyung Lee}, \bibinfo{person}{Ji~Tae
  Kim}, {and} \bibinfo{person}{Leah Lizarondo}.}
  \bibinfo{year}{2017}\natexlab{}.
\newblock \showarticletitle{A Human-Centered Approach to Algorithmic Services:
  Considerations for Fair and Motivating Smart Community Service Management
  That Allocates Donations to Non-Profit Organizations}. In
  \bibinfo{booktitle}{\emph{Proceedings of the 2017 CHI Conference on Human
  Factors in Computing Systems}} (Denver, Colorado, USA)
  \emph{(\bibinfo{series}{CHI '17})}. \bibinfo{publisher}{Association for
  Computing Machinery}, \bibinfo{address}{New York, NY, USA},
  \bibinfo{pages}{3365–3376}.
\newblock
\showISBNx{9781450346559}
\urldef\tempurl%
\url{https://doi.org/10.1145/3025453.3025884}
\showDOI{\tempurl}


\bibitem[\protect\citeauthoryear{Manders-Huits}{Manders-Huits}{2011}]%
        {Mandershuits2011}
\bibfield{author}{\bibinfo{person}{Noëmi Manders-Huits}.}
  \bibinfo{year}{2011}\natexlab{}.
\newblock \showarticletitle{What {Values} in {Design}? {The} {Challenge} of
  {Incorporating} {Moral} {Values} into {Design}}.
\newblock \bibinfo{journal}{\emph{Science and Engineering Ethics}}
  \bibinfo{volume}{17}, \bibinfo{number}{2} (\bibinfo{date}{June}
  \bibinfo{year}{2011}), \bibinfo{pages}{271--287}.
\newblock
\showISSN{1471-5546}
\urldef\tempurl%
\url{https://doi.org/10.1007/s11948-010-9198-2}
\showDOI{\tempurl}


\bibitem[\protect\citeauthoryear{Merriam and Associates}{Merriam and
  Associates}{2002}]%
        {qualresearch}
\bibfield{author}{\bibinfo{person}{Sharan~B Merriam} {and}
  \bibinfo{person}{Associates}.} \bibinfo{year}{2002}\natexlab{}.
\newblock \showarticletitle{Introduction to qualitative research}.
\newblock In \bibinfo{booktitle}{\emph{Qualitative research in practice:
  Examples for discussion and analysis}}. \bibinfo{publisher}{Jossey-Bass},
  \bibinfo{address}{Hoboken, NJ, USA}, \bibinfo{pages}{1--17}.
\newblock


\bibitem[\protect\citeauthoryear{Nguyen and Vohra}{Nguyen and Vohra}{2019}]%
        {Nguyen2019}
\bibfield{author}{\bibinfo{person}{Thành Nguyen} {and} \bibinfo{person}{Rakesh
  Vohra}.} \bibinfo{year}{2019}\natexlab{}.
\newblock \showarticletitle{Stable Matching with Proportionality Constraints}.
\newblock \bibinfo{journal}{\emph{Operations Research}} \bibinfo{volume}{67},
  \bibinfo{number}{6} (\bibinfo{year}{2019}), \bibinfo{pages}{1503--1519}.
\newblock
\urldef\tempurl%
\url{https://doi.org/10.1287/opre.2019.1909}
\showDOI{\tempurl}


\bibitem[\protect\citeauthoryear{Nissenbaum}{Nissenbaum}{2011}]%
        {Nissenbaum2011}
\bibfield{author}{\bibinfo{person}{Helen Nissenbaum}.}
  \bibinfo{year}{2011}\natexlab{}.
\newblock \showarticletitle{A Contextual Approach to Privacy Online}.
\newblock \bibinfo{journal}{\emph{Daedalus}} \bibinfo{volume}{140},
  \bibinfo{number}{4} (\bibinfo{year}{2011}), \bibinfo{pages}{32--48}.
\newblock


\bibitem[\protect\citeauthoryear{Noble}{Noble}{2018}]%
        {Noble2018}
\bibfield{author}{\bibinfo{person}{Safiya~Umoja Noble}.}
  \bibinfo{year}{2018}\natexlab{}.
\newblock \bibinfo{booktitle}{\emph{Algorithms of Oppression: How Search
  Engines Reinforce Racism}}.
\newblock \bibinfo{publisher}{New York University Press}, \bibinfo{address}{New
  York, NY, USA}.
\newblock


\bibitem[\protect\citeauthoryear{Obermeyer, Powers, Vogeli, and
  Mullainathan}{Obermeyer et~al\mbox{.}}{2019}]%
        {Obermeyer2019}
\bibfield{author}{\bibinfo{person}{Ziad Obermeyer}, \bibinfo{person}{Brian
  Powers}, \bibinfo{person}{Christine Vogeli}, {and} \bibinfo{person}{Sendhil
  Mullainathan}.} \bibinfo{year}{2019}\natexlab{}.
\newblock \showarticletitle{Dissecting racial bias in an algorithm used to
  manage the health of populations}.
\newblock \bibinfo{journal}{\emph{Science}} \bibinfo{volume}{366},
  \bibinfo{number}{6464} (\bibinfo{year}{2019}), \bibinfo{pages}{447--453}.
\newblock
\showISSN{0036-8075}
\urldef\tempurl%
\url{https://doi.org/10.1126/science.aax2342}
\showDOI{\tempurl}


\bibitem[\protect\citeauthoryear{Oosterbeek, Sóvágó, and Klaauw}{Oosterbeek
  et~al\mbox{.}}{2019}]%
        {Oosterbeek2019}
\bibfield{author}{\bibinfo{person}{Hessel Oosterbeek}, \bibinfo{person}{Sándor
  Sóvágó}, {and} \bibinfo{person}{Bas Klaauw}.}
  \bibinfo{year}{2019}\natexlab{}.
\newblock \bibinfo{booktitle}{\emph{Why are Schools Segregated? Evidence from
  the Secondary-School Match in Amsterdam}}.
\newblock \bibinfo{type}{Discussion Paper} DP13462.
  \bibinfo{institution}{Centre for Economic Policy Research}.
\newblock
\urldef\tempurl%
\url{https://ssrn.com/abstract=3319783}
\showURL{%
\tempurl}


\bibitem[\protect\citeauthoryear{Pais and Pintér}{Pais and Pintér}{2008}]%
        {Pais2008}
\bibfield{author}{\bibinfo{person}{Joana Pais} {and} \bibinfo{person}{Ágnes
  Pintér}.} \bibinfo{year}{2008}\natexlab{}.
\newblock \showarticletitle{School choice and information: An experimental
  study on matching mechanisms}.
\newblock \bibinfo{journal}{\emph{Games and Economic Behavior}}
  \bibinfo{volume}{64}, \bibinfo{number}{1} (\bibinfo{year}{2008}),
  \bibinfo{pages}{303--328}.
\newblock


\bibitem[\protect\citeauthoryear{Pathak}{Pathak}{2017}]%
        {Pathak2017}
\bibfield{author}{\bibinfo{person}{Parag~A. Pathak}.}
  \bibinfo{year}{2017}\natexlab{}.
\newblock \showarticletitle{What Really Matters in Designing School Choice
  Mechanisms}.
\newblock In \bibinfo{booktitle}{\emph{Advances in Economics and Econometrics:
  Eleventh World Congress}}, \bibfield{editor}{\bibinfo{person}{Bo~Honoré},
  \bibinfo{person}{Ariel Pakes}, \bibinfo{person}{Monika Piazzesi}, {and}
  \bibinfo{person}{Larry Samuelson}} (Eds.). \bibinfo{series}{Econometric
  Society Monographs}, Vol.~\bibinfo{volume}{1}. \bibinfo{publisher}{Cambridge
  University Press}, \bibinfo{address}{Cambridge, MA, USA},
  \bibinfo{pages}{176–214}.
\newblock
\urldef\tempurl%
\url{https://doi.org/10.1017/9781108227162.006}
\showDOI{\tempurl}


\bibitem[\protect\citeauthoryear{Pickoff-White}{Pickoff-White}{2018}]%
        {White2018}
\bibfield{author}{\bibinfo{person}{Lisa Pickoff-White}.}
  \bibinfo{year}{2018}\natexlab{}.
\newblock \bibinfo{booktitle}{\emph{S.F.'s Kindergarten Lottery: Do Parents'
  Tricks Work?}}
\newblock KQED.
\newblock
\urldef\tempurl%
\url{https://www.kqed.org/news/11641019/s-f-s-kindergarten-lottery-do-parents-tricks-work}
\showURL{%
\tempurl}


\bibitem[\protect\citeauthoryear{Rees-Jones and Skowronek}{Rees-Jones and
  Skowronek}{2018}]%
        {ReesJones2018}
\bibfield{author}{\bibinfo{person}{Alex Rees-Jones} {and}
  \bibinfo{person}{Samuel Skowronek}.} \bibinfo{year}{2018}\natexlab{}.
\newblock \showarticletitle{An experimental investigation of preference
  misrepresentation in the residency match}.
\newblock \bibinfo{journal}{\emph{Proceedings of the National Academy of
  Sciences}} \bibinfo{volume}{115}, \bibinfo{number}{45}
  (\bibinfo{year}{2018}), \bibinfo{pages}{11471--11476}.
\newblock
\showISSN{0027-8424}
\urldef\tempurl%
\url{https://doi.org/10.1073/pnas.1803212115}
\showDOI{\tempurl}


\bibitem[\protect\citeauthoryear{Roth}{Roth}{2015}]%
        {Roth2015}
\bibfield{author}{\bibinfo{person}{Alvin~E. Roth}.}
  \bibinfo{year}{2015}\natexlab{}.
\newblock \bibinfo{booktitle}{\emph{Who Gets What And Why: The Hidden World of
  Matchmaking and Market Design}}.
\newblock \bibinfo{publisher}{Houghton Mifflin Harcourt Publishing Company},
  \bibinfo{address}{New York, NY, USA}.
\newblock


\bibitem[\protect\citeauthoryear{{San Francisco Unified School District}}{{San
  Francisco Unified School District}}{2015}]%
        {sfusd2015}
\bibfield{author}{\bibinfo{person}{{San Francisco Unified School District}}.}
  \bibinfo{year}{2015}\natexlab{}.
\newblock \bibinfo{booktitle}{\emph{Student Assignment: 4th Annual Report:
  2014-15 School Year}}.
\newblock San Francisco Unified School District.
\newblock
\urldef\tempurl%
\url{https://archive.sfusd.edu/en/assets/sfusd-staff/enroll/files/2015-16/4th-annual-report-april-8-2015.pdf}
\showURL{%
Retrieved April 13, 2020 from \tempurl}


\bibitem[\protect\citeauthoryear{{San Francisco Unified School District}}{{San
  Francisco Unified School District}}{2019}]%
        {sfusd2019}
\bibfield{author}{\bibinfo{person}{{San Francisco Unified School District}}.}
  \bibinfo{year}{2019}\natexlab{}.
\newblock \bibinfo{booktitle}{\emph{Why We're Redesigning Student Assignment}}.
\newblock San Francisco Unified School District.
\newblock
\urldef\tempurl%
\url{https://www.sfusd.edu/studentassignment/why-were-redesigning-student-assignment}
\showURL{%
Retrieved April 13, 2020 from \tempurl}


\bibitem[\protect\citeauthoryear{{San Francisco Unified School District Office
  of Education}}{{San Francisco Unified School District Office of
  Education}}{nd}]%
        {policy5101}
\bibfield{author}{\bibinfo{person}{{San Francisco Unified School District
  Office of Education}}.} \bibinfo{year}{n.d.}\natexlab{}.
\newblock \bibinfo{booktitle}{\emph{Board Policy 5101: Student Assignment}}.
\newblock San Francisco Unified School District.
\newblock
\urldef\tempurl%
\url{https://go.boarddocs.com/ca/sfusd/Board.nsf/goto?open&id=B55QMC657423}
\showURL{%
Retrieved April 27, 2020 from \tempurl}


\bibitem[\protect\citeauthoryear{Saxena, Badillo-Urquiola, Wisniewski, and
  Guha}{Saxena et~al\mbox{.}}{2020}]%
        {Saxena2020}
\bibfield{author}{\bibinfo{person}{Devansh Saxena}, \bibinfo{person}{Karla
  Badillo-Urquiola}, \bibinfo{person}{Pamela~J. Wisniewski}, {and}
  \bibinfo{person}{Shion Guha}.} \bibinfo{year}{2020}\natexlab{}.
\newblock \showarticletitle{A Human-Centered Review of Algorithms Used within
  the U.S. Child Welfare System}. In \bibinfo{booktitle}{\emph{Proceedings of
  the 2020 CHI Conference on Human Factors in Computing Systems}} (Honolulu,
  HI, USA) \emph{(\bibinfo{series}{CHI ’20})}.
  \bibinfo{publisher}{Association for Computing Machinery},
  \bibinfo{address}{New York, NY, USA}, \bibinfo{pages}{1–15}.
\newblock
\showISBNx{9781450367080}
\urldef\tempurl%
\url{https://doi.org/10.1145/3313831.3376229}
\showDOI{\tempurl}


\bibitem[\protect\citeauthoryear{Scott}{Scott}{2013}]%
        {Scott2013}
\bibfield{author}{\bibinfo{person}{Janelle Scott}.}
  \bibinfo{year}{2013}\natexlab{}.
\newblock \showarticletitle{School Choice and the Empowerment Imperative}.
\newblock \bibinfo{journal}{\emph{Peabody Journal of Education}}
  \bibinfo{volume}{88}, \bibinfo{number}{1} (\bibinfo{year}{2013}),
  \bibinfo{pages}{60--73}.
\newblock
\urldef\tempurl%
\url{https://doi.org/10.1080/0161956X.2013.752635}
\showDOI{\tempurl}
\showeprint{https://doi.org/10.1080/0161956X.2013.752635}


\bibitem[\protect\citeauthoryear{Scott}{Scott}{2011}]%
        {Scott2011a}
\bibfield{author}{\bibinfo{person}{Janelle~T. Scott}.}
  \bibinfo{year}{2011}\natexlab{}.
\newblock \showarticletitle{Market-Driven Education Reform and the Racial
  Politics of Advocacy}.
\newblock \bibinfo{journal}{\emph{Peabody Journal of Education}}
  \bibinfo{volume}{86}, \bibinfo{number}{5} (\bibinfo{year}{2011}),
  \bibinfo{pages}{580--599}.
\newblock
\urldef\tempurl%
\url{https://doi.org/10.1080/0161956X.2011.616445}
\showDOI{\tempurl}


\bibitem[\protect\citeauthoryear{Selbst, {b}oyd, Friedler, Venkatasubramanian,
  and Vertesi}{Selbst et~al\mbox{.}}{2019}]%
        {Selbst2019}
\bibfield{author}{\bibinfo{person}{Andrew~D. Selbst}, \bibinfo{person}{{d}anah
  {b}oyd}, \bibinfo{person}{Sorelle~A. Friedler}, \bibinfo{person}{Suresh
  Venkatasubramanian}, {and} \bibinfo{person}{Janet Vertesi}.}
  \bibinfo{year}{2019}\natexlab{}.
\newblock \showarticletitle{Fairness and Abstraction in Sociotechnical
  Systems}. In \bibinfo{booktitle}{\emph{Proceedings of the Conference on
  Fairness, Accountability, and Transparency}} (Atlanta, GA, USA)
  \emph{(\bibinfo{series}{FAT* ’19})}. \bibinfo{publisher}{Association for
  Computing Machinery}, \bibinfo{address}{New York, NY, USA},
  \bibinfo{pages}{59–68}.
\newblock
\showISBNx{9781450361255}
\urldef\tempurl%
\url{https://doi.org/10.1145/3287560.3287598}
\showDOI{\tempurl}


\bibitem[\protect\citeauthoryear{Shapley and Scarf}{Shapley and Scarf}{1974}]%
        {ShapleyScarf1974}
\bibfield{author}{\bibinfo{person}{Lloyd Shapley} {and}
  \bibinfo{person}{Herbert Scarf}.} \bibinfo{year}{1974}\natexlab{}.
\newblock \showarticletitle{On cores and indivisibility}.
\newblock \bibinfo{journal}{\emph{Journal of Mathematical Economics}}
  \bibinfo{volume}{1}, \bibinfo{number}{1} (\bibinfo{year}{1974}),
  \bibinfo{pages}{23 -- 37}.
\newblock
\showISSN{0304-4068}
\urldef\tempurl%
\url{https://doi.org/10.1016/0304-4068(74)90033-0}
\showDOI{\tempurl}


\bibitem[\protect\citeauthoryear{Shilton}{Shilton}{2013}]%
        {Shilton2013}
\bibfield{author}{\bibinfo{person}{Katie Shilton}.}
  \bibinfo{year}{2013}\natexlab{}.
\newblock \showarticletitle{Values Levers: Building Ethics into Design}.
\newblock \bibinfo{journal}{\emph{Science, Technology, \& Human Values}}
  \bibinfo{volume}{38}, \bibinfo{number}{3} (\bibinfo{year}{2013}),
  \bibinfo{pages}{374--397}.
\newblock
\urldef\tempurl%
\url{https://doi.org/10.1177/0162243912436985}
\showDOI{\tempurl}


\bibitem[\protect\citeauthoryear{Smith, Yu, Srivastava, Halfaker, Terveen, and
  Zhu}{Smith et~al\mbox{.}}{2020}]%
        {Smith2020}
\bibfield{author}{\bibinfo{person}{C.~Estelle Smith}, \bibinfo{person}{Bowen
  Yu}, \bibinfo{person}{Anjali Srivastava}, \bibinfo{person}{Aaron Halfaker},
  \bibinfo{person}{Loren Terveen}, {and} \bibinfo{person}{Haiyi Zhu}.}
  \bibinfo{year}{2020}\natexlab{}.
\newblock \showarticletitle{Keeping Community in the Loop: Understanding
  Wikipedia Stakeholder Values for Machine Learning-Based Systems}. In
  \bibinfo{booktitle}{\emph{Proceedings of the 2020 CHI Conference on Human
  Factors in Computing Systems}} (Honolulu, HI, USA)
  \emph{(\bibinfo{series}{CHI ’20})}. \bibinfo{publisher}{Association for
  Computing Machinery}, \bibinfo{address}{New York, NY, USA},
  \bibinfo{pages}{1–14}.
\newblock
\showISBNx{9781450367080}
\urldef\tempurl%
\url{https://doi.org/10.1145/3313831.3376783}
\showDOI{\tempurl}


\bibitem[\protect\citeauthoryear{Sweeney}{Sweeney}{2013}]%
        {Sweeney2013}
\bibfield{author}{\bibinfo{person}{Latanya Sweeney}.}
  \bibinfo{year}{2013}\natexlab{}.
\newblock \showarticletitle{Discrimination in Online Ad Delivery}.
\newblock \bibinfo{journal}{\emph{Commun. ACM}} \bibinfo{volume}{56},
  \bibinfo{number}{5} (\bibinfo{date}{May} \bibinfo{year}{2013}),
  \bibinfo{pages}{44–54}.
\newblock
\showISSN{0001-0782}
\urldef\tempurl%
\url{https://doi.org/10.1145/2447976.2447990}
\showDOI{\tempurl}


\bibitem[\protect\citeauthoryear{Voida, Dombrowski, Hayes, and Mazmanian}{Voida
  et~al\mbox{.}}{2014}]%
        {Voida2014}
\bibfield{author}{\bibinfo{person}{Amy Voida}, \bibinfo{person}{Lynn
  Dombrowski}, \bibinfo{person}{Gillian~R. Hayes}, {and}
  \bibinfo{person}{Melissa Mazmanian}.} \bibinfo{year}{2014}\natexlab{}.
\newblock \showarticletitle{Shared Values/Conflicting Logics: Working around
  e-Government Systems}. In \bibinfo{booktitle}{\emph{Proceedings of the SIGCHI
  Conference on Human Factors in Computing Systems}} (Toronto, Ontario, Canada)
  \emph{(\bibinfo{series}{CHI '14})}. \bibinfo{publisher}{Association for
  Computing Machinery}, \bibinfo{address}{New York, NY, USA},
  \bibinfo{pages}{3583–3592}.
\newblock
\showISBNx{9781450324731}
\urldef\tempurl%
\url{https://doi.org/10.1145/2556288.2556971}
\showDOI{\tempurl}


\bibitem[\protect\citeauthoryear{Zhu, Yu, Halfaker, and Terveen}{Zhu
  et~al\mbox{.}}{2018}]%
        {Zhu2018}
\bibfield{author}{\bibinfo{person}{Haiyi Zhu}, \bibinfo{person}{Bowen Yu},
  \bibinfo{person}{Aaron Halfaker}, {and} \bibinfo{person}{Loren Terveen}.}
  \bibinfo{year}{2018}\natexlab{}.
\newblock \showarticletitle{Value-Sensitive Algorithm Design: Method, Case
  Study, and Lessons}.
\newblock \bibinfo{journal}{\emph{Proc. ACM Hum.-Comput. Interact.}}
  \bibinfo{volume}{2}, \bibinfo{number}{CSCW}, Article \bibinfo{articleno}{194}
  (\bibinfo{date}{Nov.} \bibinfo{year}{2018}), \bibinfo{numpages}{23}~pages.
\newblock
\urldef\tempurl%
\url{https://doi.org/10.1145/3274463}
\showDOI{\tempurl}


\end{thebibliography}

\appendix

\end{document}